\documentclass[fleqn,usenatbib]{mnras}

\usepackage{newtxtext,newtxmath}

\usepackage[T1]{fontenc}
\usepackage{float}

\DeclareRobustCommand{\VAN}[3]{#2}
\let\VANthebibliography\thebibliography
\def\thebibliography{\DeclareRobustCommand{\VAN}[3]{##3}\VANthebibliography}

\usepackage{graphicx}	
\usepackage{amsmath}	

\usepackage{amssymb}	

\usepackage{booktabs} 
\usepackage{siunitx}  

\newcommand{\simname}[1]{{\fontfamily{lmss}\fontseries{sbc}\selectfont #1}}

\title[IMBHs and NSCs in merging low-mass galaxies] {The coupled evolution of nuclear star clusters and intermediate-mass black holes in a dwarf-galaxy merger}

\author[C. Partmann et al.]{Christian Partmann$^{1, 2}$\thanks{E-mail: christian.partmann@epfl.ch},
Rachel S. Somerville$^{2}$,
Michaela Hirschmann$^{1}$,
Antti Rantala$^{3,4}$,
Natalia Lah\'en$^{5}$
\newauthor
\\
$^{1}$Institute of Physics, Laboratory for Galaxy Evolution and Spectral Modelling, École Polytechnique Fédérale de Lausanne, \\
Observatoire de Sauverny, Chemin Pegasi 51, CH-1290 Versoix, Switzerland\\
$^{2}$Center for Computational Astrophysics, Flatiron Institute, 162 5th Avenue, New York, NY 10010, USA\\
$^{3}$Institute of Astronomy, University of Cambridge, Madingley Road, Cambridge CB3 0HA, UK\\
$^{4}$Kavli Institute for Cosmology, Cambridge (KICC), University of Cambridge, Madingley Road, Cambridge CB3 0HA, UK\\
$^{5}$Max-Planck-Institut für Astrophysik, Karl-Schwarzschild-Str. 1, D-85748 Garching, Germany\\
}

\date{Accepted XXX. Received YYY; in original form ZZZ}

\pubyear{2026}

\begin{document}
\label{firstpage}
\pagerange{\pageref{firstpage}--\pageref{lastpage}}
\maketitle

\begin{abstract}
Nuclear star clusters (NSCs) and intermediate-mass black holes (IMBHs) are expected to coexist in low-mass galaxies and might together facilitate the coalescence and growth of IMBHs. We simulate an equal-mass, gas-rich dwarf-galaxy merger in which each progenitor hosts a $10^3\,\rm M_\odot$ IMBH embedded in a $10^6\,\rm M_\odot$ NSC to study their co-evolution. The simulation model combines the multiphase interstellar medium and fully sampled initial stellar mass function of {\sc griffin} at $4\,\rm M_\odot$ gas resolution with the regularised post-Newtonian black hole dynamics of {\sc Ketju}. Black holes are modelled as accreting sink particles without feedback. Without NSCs, $10^3$--$10^4\,\rm M_\odot$ IMBHs sink only to separations of $50$--$200\,\rm pc$ and capture no gas. With NSCs, the clusters coalesce within ${\sim} 500\,\rm Myr$, the IMBHs form a binary within a few Myr, and scattering of mainly old NSC stars drives them to merger ${\sim} 340\,\rm Myr$ later. The NSCs also promote gas capture, growing the black holes to $1.1\times10^5$ and $4.1\times10^4\,\rm M_\odot$. Gas capture is suppressed during the merger-induced starburst by stellar feedback and the motion of the NSCs relative to the gas, and proceeds in brief supernova-regulated episodes once the NSCs have settled in the centre.  Scattering by the IMBH binary and the merger recoil of ${\sim} 175\,\rm km\,s^{-1}$, which removes the remnant from the NSC but not from the galaxy, leave a larger, rotating NSC with a flat core. The stellar tidal disruption rate rises towards coalescence, and the merger is detectable with LISA out to high redshift. NSCs thus remove the delivery and hardening bottlenecks of IMBH mergers in low-mass galaxies.
\end{abstract}




\section{Introduction}
Intermediate-mass black holes (IMBHs, $M_{\rm BH} \sim 10^{2}$--$10^{5}\,\rm M_\odot$) bridge the gap between stellar-mass black holes (BHs) and supermassive black holes \citep[SMBHs with $M_{\rm BH} > 10^{6}\,\rm M_\odot$, see][for a review]{Greene_2020}. While their origin is still debated, they are the missing link to understand where the SMBHs that reside at the centres of most massive galaxies form.

Although they are difficult to detect, growing observational samples provide evidence for accreting IMBHs and BHs just above the IMBH mass range (${\sim} 10^5$--$10^6 \, \rm M_\odot$), in dwarf galaxies \citep{Filippenko_Ho_2003, Baldassare_2015}, including more than 150 optical candidates in the SDSS \citep{Reines_2013}, X-ray-selected IMBH candidates out to $z\sim2.4$ \citep{Mezcua_2016,Mezcua_2018}, X-ray selected active galactic nuclei (AGN) in local dwarfs from eROSITA \citep{Bykov_2024,Sacchi_2024}, and dynamical measurements \citep{Nguyen_2019}. Recently, DESI spectroscopy has tripled the census of AGN candidates in dwarf galaxies, including several hundred IMBH candidates \citep{Pucha_2025}. One of the most robust IMBH candidates resides in the star cluster $\omega$~Centauri and has a dynamically measured mass of $\gtrsim8200\,\rm M_\odot$ \citep{Haberle_2024}. Nevertheless, the occupation fraction and mass distribution of IMBHs, particularly in low-mass galaxies that are expected to host this BH population, remain uncertain.

IMBHs might also be important as progenitors of SMBHs observed today. Indeed, JWST has uncovered a vast population of massive black holes at high redshift \citep{Ubler_2023,Harikane_2023,Kocevski_2023,Matthee_2024}, with some candidates close to the IMBH range \citep{Maiolino_2024b,Fei_2026}. However, most BHs with masses below $10^5 \, \rm M_\odot$ are likely still missed by current observations. While some IMBHs at high redshift are expected to grow into SMBHs, some of them will be in environments where IMBHs do not grow efficiently. Hence, studying their evolution might help to understand what information their demographics retain about the formation of the first BHs \citep{Greene_2020,Volonteri_2021}.

IMBHs can grow via gas accretion (including the disruption of stars) or mergers with other BHs. Especially at high redshift, mergers between galaxies are frequent \citep{Fakhouri_2010,Deason_2014}, but for BHs to grow via galaxy mergers the BHs brought into the galaxy must coalesce. Whether this process is efficient for IMBHs is therefore central to models of hierarchical black hole growth.

Coalescence proceeds through dynamical-friction-driven sinking, stellar hardening, and finally gravitational-wave (GW) emission \citep{Peters_1964,Begelman_1980,Quinlan_1996}. Each stage can be inefficient for IMBHs. As dynamical friction depends on the mass of the sinking body \citep{Chandrasekhar_1943}, pairs in low-mass haloes and gas-poor galaxies may remain off-centre for more than a Hubble time \citep{Tremmel_2018,Bellovary_2019,Pfister_2019,Bortolas_2020,Bellovary_2021,Ma_2021,Chen_2024,2024MNRAS.532.4681P,Zhou_2025}, particularly in the cored density profiles common in dwarfs \citep{Read_2006,Tamfal_2018}. Even after a bound binary forms, stellar scattering must remove orbital energy from the binary for its separation to shrink (also called ``stellar hardening''). However, the scattering of stars can be inefficient outside of dense star clusters and the binary can stall once it depletes the stellar orbits available for three-body scattering \citep{Milosavljevic_2003,Khan_2024}. In this case, more complex interactions, e.g. with a third BH, are required to trigger a BH merger \citep{2024MNRAS.532.4681P}. Finally, as the \citet{Schwarzschild_1916} radius is BH mass dependent, the separation of IMBHs must be smaller than for SMBHs for gravitational waves to drive the BHs to merge \citep{Peters_1964}.

Nuclear star clusters (NSCs) may alleviate both the sinking and hardening problems, by enhancing the mass of the sinking body and by providing high stellar densities for the scattering of stars.  Direct $N$-body simulations find that IMBH binaries embedded in sufficiently massive NSCs can form hard binaries and merge within a few hundred Myr \citep{Khan_2021,Khan_2026}. When each black hole is initially carried by its own NSC, the interaction and merger of the clusters can further accelerate the early orbital decay \citep[][see also \citealt{MastrobuonoBattisti_2025}]{Ogiya_2020,Mukherjee_2025}. NSCs are common in low- and intermediate-mass galaxies, with nucleation fractions peaking around $M_\star\sim10^9\,\rm M_\odot$ \citep{SanchezJanssen_2019,Neumayer_2020,Hoyer_2021}, although the fraction that also hosts massive black holes remains poorly constrained due to observational limitations \citep{Hoyer_2024}. Dense clusters also provide a plausible scenario for the formation of IMBH seeds through runaway stellar collisions, tidal disruption of stars, and hierarchical black hole mergers \citep{PortegiesZwart_2002,Giersz_2015,Stone_2017,Antonini_2019,Rizzuto_2021,Fujii_2024,Rantala_2024}. If IMBH seeds indeed form in such clusters, the IMBHs delivered by dwarf-galaxy mergers arrive embedded in exactly the stellar systems that can both carry them to the centre and harden their binary, which motivates studying the two together. In addition, their potential well can enhance the mass growth of IMBHs via gas accretion \citep{2025MNRAS.537..956P, 2025MNRAS.544.2467S}, allowing even IMBHs of ${\sim} 100 \, \rm M_\odot$ to grow fast \citep{2025MNRAS.537..956P}. BHs in NSCs are also predicted to be the dominant source of tidal disruption events (TDEs), which  are relevant both as a BH growth channel and as a tool to identify IMBHs observationally \citep{Polkas2024}.

Galaxy mergers can also reshape NSCs: they may grow in mass via coalescence, be tidally stripped, or acquire young stars formed from captured gas. These outcomes connect to the two main NSC formation channels, cluster inspiral and in-situ nuclear star formation \citep{Tremaine_1975,Milosavljevic_2004,Neumayer_2020}. In dwarf galaxies, studies indicate that cluster inspiral dominates NSC growth, so that their NSCs generally contain old, metal-poor populations with smaller young components \citep{Fahrion_2021,Fahrion_2022}. Ongoing cluster mergers observed in several dwarf nuclei provide direct evidence for this assembly channel \citep{Poulain_2025}. Furthermore, a BH binary produced in the merger of NSCs during a galaxy merger can affect the structure of the NSC: first, via the slingshot ejection of stars with high velocity during the stellar hardening phase \citep{Quinlan_1996}, and second, if the BH merger produces a large recoil, through the ejection of the merger remnant, which unbinds stars in the centre of the NSC \citep{Blecha_2008,Gualandris_2008}.

Gas-rich mergers may simultaneously promote IMBH growth. Galaxy interactions have been shown to enhance black hole activity in massive galaxies \citep[observationally and in simulations, respectively]{Ellison_2011,Jhee_2026}, while dwarf pairs show enhanced star formation \citep{Stierwalt_2015}, and can undergo nuclear starbursts that form massive bound clusters \citep{2019ApJ...879L..18L,2020ApJ...891....2L,Zhang_2020a,Sun_2025}. Observational evidence for black hole fuelling in dwarf mergers is also emerging: candidate dual AGN \citep{Micic_2023,Mezcua_2026} and a luminous AGN in a late-stage dwarf merger \citep{Kimbrell_2024} have been identified, and dwarf pairs show a higher AGN fraction than isolated dwarfs \citep{Micic_2024,Erostegui_2025}. These results motivate studying IMBH dynamics and gas accretion together rather than as separate processes.

The state-of-the-art simulations in the literature capture only parts of this coupled problem. Large-scale cosmological simulations generally reposition black holes to the potential minimum of the host galaxy \citep{SpringelDiMatteoHernquist_2005,Sijacki_2007} and do not even attempt to resolve BH dynamics, while recent simulation suites have invoked subgrid models based on \citet{Chandrasekhar_1943} to account for unresolved dynamical-friction forces \citep{Hirschmann_2014,Steinborn_2015,Tremmel_2015,Tremmel_2018,Bellovary_2019,Bellovary_2021,Genina2024,Bhowmick_2025,Li_2025,Bhowmick_2026}. Dynamical friction subgrid models allow for physically motivated BH sinking to the centres of galaxies, but cannot resolve scattering by the nuclear stellar population. Going one step further, models have been proposed that follow the unresolved evolution of BH binaries with a subgrid prescription, complementary to simulations that resolve the small-scale dynamics explicitly \citep{Li_2025}. Collisional $N$-body calculations can resolve the complex internal NSC and BH binary dynamics but usually omit gas, star formation, and feedback \citep{Ogiya_2020, Khan_2021,Mukherjee_2025,Souvaitzis_2025, Khan_2026}. Conversely, simulations with an individually sampled stellar population and a resolved multiphase interstellar medium (ISM) have so far concentrated on single BHs in isolated galaxies with and without nuclear star clusters \citep{2025MNRAS.537..956P,2025A&A...704A.177P,2025MNRAS.544.2467S,Shin2026}.

One approach to handle the range of scales involved in the BH coalescence process is to use regularised integration schemes like {\sc Ketju} \citep{Rantala_2017,2020MNRAS.492.4131R,Mannerkoski_2023} to resolve the interaction of BHs with their stellar environment, an approach that has been used for mergers of massive galaxies \citep{Liao_2024a,Liao_2024b,Wright_2026} and cosmological zoom-in simulations \citep{2021ApJ...912L..20M,Keitaanranta2026}. In particular, the study by \citet{Liao_2024a} points out the importance of nuclear star formation for the coalescence of BHs in massive galaxies, but without sufficient spatial resolution to resolve a nuclear star cluster. A promising approach was presented by the MAGICS project \citep{Chen_2024, Zhou_2025}, which re-simulated galaxy interactions from a large-scale cosmological simulation with increasing detail and found that BHs stripped of their galactic stellar environment often stall before forming BH binaries. By re-simulating these galaxy interactions with a collisional $N$-body code and adding an NSC by hand, \citet{Mukherjee_2025} showed that the sinking and formation of BH binaries can be accelerated by the presence of an NSC. With this approach, \citeauthor{Mukherjee_2025} followed the formation and evolution of IMBH binaries from galactic scales to the onset of the binary hardening phase, but without accounting for gas and without following the hardening phase to the gravitational-wave-driven merger.

In this work, we overcome some limitations of previous studies by combining the {\sc griffin} model \citep{Hu_2014,Hu_2016,Hu_2017, 2020ApJ...891....2L, 2025MNRAS.537..956P} with the regularised post-Newtonian black hole dynamics of {\sc Ketju} \citep{Rantala_2017,2020MNRAS.492.4131R,Mannerkoski_2023}. {\sc griffin} provides a resolved multiphase ISM and a star formation recipe that samples individual stars over the whole initial mass function (IMF) and thereby produces realistic star cluster properties, while {\sc Ketju} can be used to resolve BH sinking, stellar scattering, and the GW-driven merger. We simulate the merger of two equal-mass dwarf galaxies, each hosting a live NSC with an IMBH at its centre. This set-up allows us to follow explicitly, to our knowledge for the first time, the sinking, gas capture, nuclear star formation, NSC merger, BH binary hardening, and gravitational-wave-driven coalescence of two IMBHs through a gas-rich dwarf-galaxy merger.

The paper is structured as follows. In Section~\ref{sec:method}, we summarise the galaxy model, the sink-particle-based BH accretion model, and the regularised integrator {\sc Ketju}. The initial conditions are described in Section~\ref{sec:IC}. Section~\ref{sec:results} presents our results, beginning with an overview of the fiducial simulation. We then examine BH sinking and growth through gas accretion in Section~\ref{subsec:growth}, the feedback processes regulating BH growth in Section~\ref{sec:fbregulation}, and the BH merger and recoil in Sections~\ref{sec:Coalesence} and \ref{sec:merger_recoil}. The evolution of the NSC is discussed separately in Section~\ref{sec:NSCevo}. Potentially observable signatures of this evolution -- the structure and kinematics of the NSC, tidal disruption events and the gravitational-wave signal -- are presented in Section~\ref{sec:discussion_observables}. We discuss our results in Section~\ref{sec:Discussion} and summarise our conclusions in Section~\ref{Sec:summary}.

\begin{table*}
\centering
\caption{Structural and numerical parameters of each dwarf galaxy, following, in modified form, \citet{Hu_2016,Hu_2017} and \citet{2025MNRAS.537..956P}. $m_{\rm res}$, $\epsilon_{\rm soft}$ and $N_{\rm part}$ are the particle mass, gravitational softening length and particle number of each component. The NSC is only present in \simname{1e3-NSC},  \simname{1e3-NSC-isolated} and \simname{1e3-NSC-reset} (Table~\ref{tab:simulations}). }
\label{tab:ics}
\begin{tabular}{l c c c c c c c}
\toprule
Component & $M\,[\rm M_\odot]$ & size [kpc] & profile & $m_{\rm res}\,[\rm M_\odot]$ & $\epsilon_{\rm soft}$ [pc] & $N_{\rm part}$ & interaction with BH\\
\midrule
Dark matter halo & $2\times10^{10}$ & 44 ($r_{\rm vir}$) & Hernquist ($\lambda=0.03$) & 400 & 1.0 & $6.4\times10^{7}$ & softened\\
Gas disc & $4\times10^{7}$ & 1.46 ($r_{\rm gas}$) & exponential & 4.0  & 0.1 & $1.0\times10^{7}$ & softened\\
Stellar disc & $2\times10^{7}$ & 0.73 ($r_*$) & exponential ($h_* = 0.35\,\rm kpc$)& 4.0 & 0.05 & $5.0\times10^{6}$ & unsoftened\\
old NSC & $1\times10^{6}$ & $3\times10^{-3}$ ($r_{1/2}$) & Dehnen ($\gamma=1.75$) & 20.0 & 0.05 & $5.0\times10^{4}$ & unsoftened for $t > t_{\rm Ketju}$\\
New stars & - & - & - & full IMF > 0.08 & 0.05 & - & unsoftened \\
\bottomrule
\end{tabular}
\end{table*}

\section{Method}
\label{sec:method}
\subsection{Galaxy evolution model}
\label{griffinmodel}
The simulations employ the {\sc griffin} simulation model \citep{Hu_2014,Hu_2016,Hu_2017, 2020ApJ...891....2L}, which is designed to model the multiphase interstellar medium at solar mass resolution in a galactic context and includes non-equilibrium thermochemistry, radiative heating and cooling, star formation of individual stars covering the whole initial mass function and stellar feedback from individual massive stars. The details of the model, which has been used and improved extensively in various studies of isolated and interacting dwarf galaxies \citep[e.g.][]{2019MNRAS.483.3363H,2020ApJ...891....2L,Steinwandel_2020,2022MNRAS.509.5938H,Lah_n_2023,Lahen_2025a,Lahen_2025b,2025MNRAS.537..956P}, are described in \citet{Hu_2014,Hu_2016,Hu_2017, Lah_n_2023}, and we give a brief summary below.

The {\sc griffin} model is implemented in {\sc sphgal} \citep{Hu_2014}, an extension of the smoothed particle hydrodynamics (SPH) code {\sc gadget-3} \citep{Springel_2005}, with a pressure--energy based formulation of the equations of hydrodynamics \citep{ediss19451} using a Wendland C4 kernel with $N_{\rm ngb}=100$ neighbours \citep{2012MNRAS.425.1068D}.

Following \citet{Lah_n_2023}, star formation is modelled with a Jeans limit, which converts gas into hydrodynamically decoupled reservoir particles as soon as the Jeans mass falls below half the SPH kernel mass. After one local dynamical time, $t_{\rm dyn}=(4\pi G\rho)^{-1/2}$, reservoirs stochastically form individual stars drawn from the \citet{2001MNRAS.322..231K} IMF over $0.08$--$500\,\rm M_\odot$. Stellar lifetimes are taken from the Geneva models at $Z=0.1\,\rm Z_\odot$ \citep{2013A&A...558A.103G}; small random offsets in stellar position and momentum at birth account for unresolved substructure \citep[see][for details]{Lah_n_2023}.

Gas between $10 \,\rm K$ and $3\times10^4\,\rm K$ is evolved with a non-equilibrium chemical network for $\rm H_2$, $\rm H$, $\rm H^+$, CO, $\rm C^+$, O, and free electrons \citep{1997ApJ...482..796N,2007ApJS..169..239G,2012MNRAS.421..116G,Hu_2016}, adopting a dust-to-gas ratio of $0.1$ per cent. At higher temperatures, metal-dependent equilibrium cooling tables are used \citep{2009MNRAS.393...99W}. Stellar far-ultraviolet (FUV) and ionising luminosities are obtained from the {\sc BaSeL} library together with the Geneva tracks \citep{1997A&AS..125..229L,1998A&AS..130...65L,2002A&A...381..524W,2013A&A...558A.103G}. FUV radiation from stars within $50\,\rm pc$ is attenuated using columns computed with {\sc TreeCol} \citep{Clark_2011}, while ionising sources above $8\,\rm M_\odot$ are treated with a Str\"omgren-sphere approximation that heats ionised gas to $10^4\,\rm K$ \citep{2012MNRAS.421.3488H,Hu_2017}. Asymptotic giant branch (AGB) feedback from $0.5$--$8\,\rm M_\odot$ stars is injected at the end of their lifetimes \citep{2010MNRAS.403.1413K}. Stars with initial masses of $8$--$50\,\rm M_\odot$ undergo core-collapse supernovae (SNe) and inject $10^{51}\,\rm erg$ of thermal energy together with metal yields \citep{2004ApJ...608..405C}. The ejecta are distributed approximately isotropically with a 12-bin {\sc healpix} tessellation \citep{Gorski_2005}, depositing material into the nearest $8\pm2$ gas particles per angular bin \citep{2019MNRAS.483.3363H}. Because the mass resolution is $4\,\rm M_\odot$, the thermal injection captures the adiabatic expansion of individual remnants without requiring a subgrid correction for numerical overcooling \citep{Hu_2016, Steinwandel_2020}. Stars initially more massive than $50\,\rm M_\odot$ are assumed to collapse directly to stellar-mass black holes without an SN when they reach the end of their lifetimes.

\begin{table}
\centering
\caption{Summary of simulations. Aside from the fiducial simulation \simname{1e3-NSC}, we present three simulations without NSCs for comparison, a simulation of one of the progenitor galaxies of \simname{1e3-NSC} in isolation, and a numerical experiment in which the BH masses are reset to $10^3\,\rm M_\odot$ at $t=460\,\rm Myr$ (Section~\ref{sec:Coalesence}).}
\label{tab:simulations}
\begin{tabular}{l S[table-format=2.0] S[table-format=7.0] l }
\toprule
Simulation name & {Initial BH mass} & {NSC mass} & IC \\
\midrule
\simname{1e3-NSC} (fiducial)    & \num{1e3} $\rm M_\odot$ & \num{1e6} $\rm M_\odot$& merger \\
\simname{1e3-noNSC}   & \num{1e3} $\rm M_\odot$ & \text{--} & merger \\
\simname{1e4-noNSC}   & \num{1e4} $\rm M_\odot$ & \text{--} & merger \\
\simname{1e6-noNSC}   & \num{1e6} $\rm M_\odot$ & \text{--} & merger \\
\midrule
\simname{1e3-NSC-isolated} & \num{1e3} $\rm M_\odot$ & \num{1e6} $\rm M_\odot$& isolated \\
\midrule
{\simname{1e3-NSC-reset}} & {\num{1e3} $\rm M_\odot$} & {\num{1e6} $\rm M_\odot$} & merger \\
\bottomrule
\end{tabular}
\end{table}

\subsection{BH accretion model}
\label{sec:BHaccretion}
IMBHs are represented by sink particles that accrete gas following the prescription of \citet{Bate_1995}, as implemented and validated in {\sc griffin} by \citet{2025MNRAS.537..956P}. A gas particle within a fixed accretion radius $r_{\rm acc}$ is accreted onto a BH of mass $M_{\rm BH}$ if it is gravitationally bound, $\frac{1}{2}\lvert \boldsymbol{v} \rvert^2 + u_{\rm gas} < \frac{G M_{\rm BH}}{\lvert \boldsymbol{r} \rvert}$, and its specific angular momentum can no longer sustain a circular orbit at $r_{\rm acc}$, i.e.~$l = \lvert \boldsymbol{r} \times \boldsymbol{v} \rvert < \sqrt{G M_{\rm BH} r_{\rm acc}}$, where $\boldsymbol{r}$ and $\boldsymbol{v}$ are the position and velocity of the gas particle relative to the BH. We adopt $r_{\rm acc} = 0.5\,\rm pc$, comparable to the SPH smoothing length of star-forming gas and well above the employed gravitational softening of the gas of $0.1 \, \rm pc$ \citep[see][for a resolution test]{2025MNRAS.537..956P}. Accreted SPH particles are added to the dynamical mass of the BH in a mass and momentum conserving fashion. To reduce the model complexity, we do not include BH feedback in this study.

The model resolves the typical Bondi--Hoyle--Lyttleton radius of the BHs in our simulations (e.g.\ $r_{\rm Bondi}\sim0.9-30\,\rm pc$ for a $10^4\,\rm M_\odot$ BH across the warm-to-cold ISM phases), but not the accretion disc itself, about which we do not make any additional assumptions. Together with the ISM model described in the previous Section~\ref{griffinmodel}, this prescription allows us to simulate the bursty BH accretion history and the evolution of the nuclear gas supply, which feeds the BH and leads to nuclear star formation \citep[see][for details]{2025MNRAS.537..956P}.

\subsection{BH dynamics model: Ketju}

We model the dynamics of the IMBHs using the regularised gravity scheme {\sc Ketju} \citep{Rantala_2017,2018ApJ...864..113R,2021ApJ...912L..20M,Mannerkoski_2023}, which extends the {\sc Gadget-3} tree code with a high-accuracy treatment of close gravitational encounters. Around each BH, a spherical ``Ketju region'' of radius $r_{\rm Ketju}$ is defined. Inside this region, the standard leapfrog integration is replaced by the {\sc MSTAR} algorithmic regularisation method \citep{2020MNRAS.492.4131R}. {\sc MSTAR} employs a time-transformed formulation of the equations of motion together with a minimum-spanning-tree coordinate representation and Gragg--Bulirsch--Stoer (GBS) extrapolation \citep{Gragg_1965,Bulirsch_1966}, enabling close BH--star and BH--BH encounters to be integrated without gravitational softening at essentially machine precision. In contrast, forces between stars are generally softened in this study. Particles can enter and leave the regularised region continuously, provided that $r_{\rm Ketju} > 2.8\,\epsilon_{\rm soft}$. This hybrid treatment allows us to follow both the dynamical-friction-driven orbital decay of the IMBHs and, after a bound binary forms, its subsequent hardening through interactions with the surrounding stellar population.

The mutual BH--BH interaction is integrated with post-Newtonian corrections up to order $3.5$, capturing the gravitational-wave-driven inspiral. In this project, two BHs are merged once their semimajor axis falls below $100$ Schwarzschild radii ($R_{\rm S}$), and the merger remnant is assigned a recoil kick based on the mass ratio and spins of the merging pair, following the fitting formulae of \citet{2015PhRvD..92b4022Z}. We use a GBS integration tolerance of $\eta = 10^{-7}$, which \citet{Mannerkoski_2023} find sufficient to reproduce the orbital evolution of an isolated post-Newtonian binary to better than 1 per cent. The size of the Ketju region is $r_{\rm Ketju} = 3\,\epsilon_{\rm soft,*} = 0.15 \, \rm pc$. Details on how different particle types are treated are given in the next Section~\ref{sec:IC} and Table~\ref{tab:ics}.

\begin{figure*}
	\includegraphics[width=2\columnwidth]{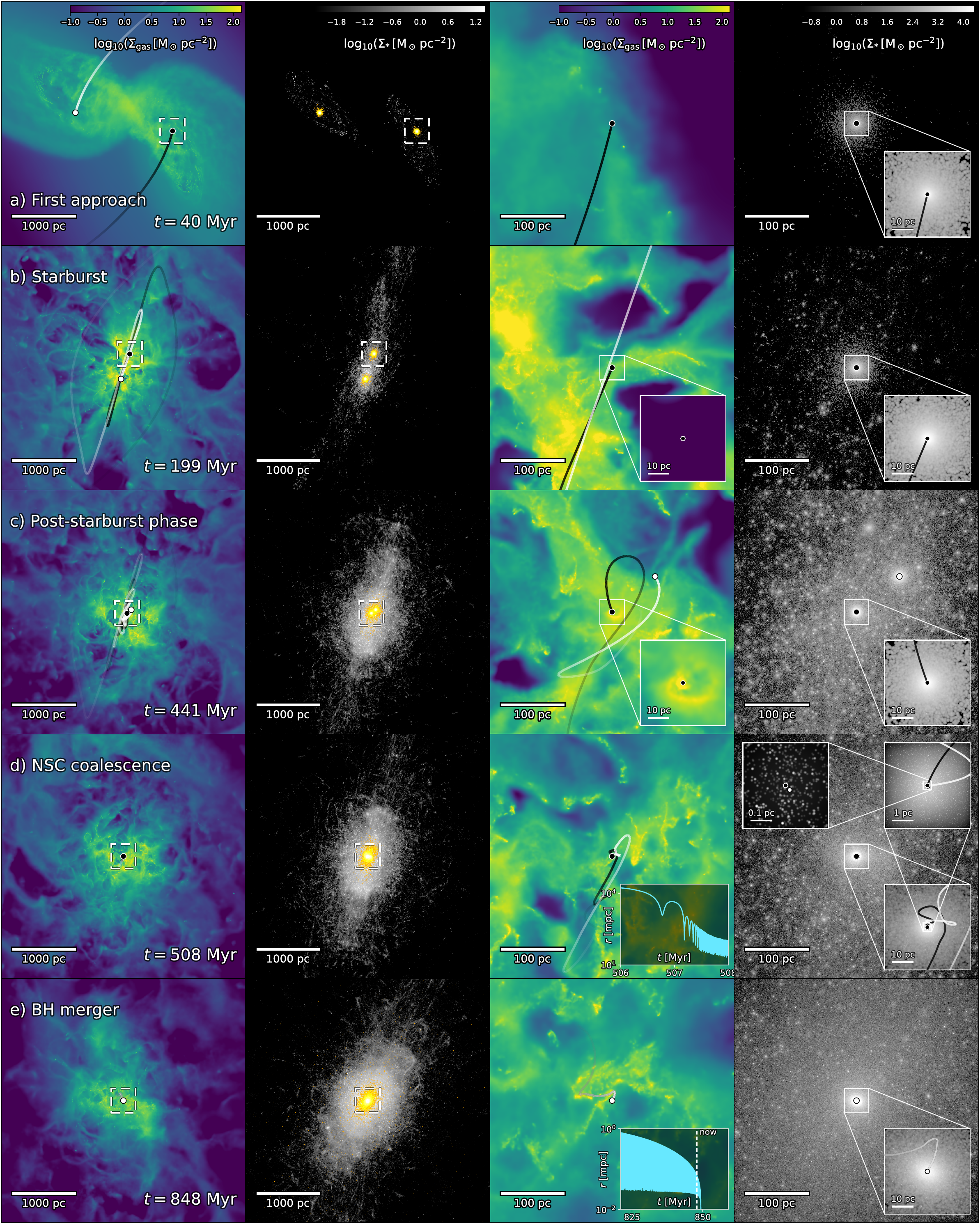}
 \caption{Overview of the evolution of the fiducial merger simulation \simname{1e3-NSC} at (a) the time of first approach, (b) the starburst triggered by the galaxy interaction, (c) the post-starburst phase during which the BHs and NSCs are still unbound, (d) the NSC coalescence and BH binary formation, and (e) the BH merger phase. The first and second columns show the gas and stellar surface densities on galactic scales (scale bar: $1000\,\rm pc$), showing the full extent of both interacting galactic discs. To guide the eye, we over-plot the stellar surface density of new stars with the pre-existing NSCs in yellow/red. The third and fourth columns show gas and stellar surface densities zoomed in on the region around BH-1 (scale bar: $100\,\rm pc$). The trajectories of the two black holes are shown in black and white, respectively. Zoom-ins on the region around the BH show the gas supply (panels b and c, column 3) and the stellar environment of the BH (column 4). The evolution of the BH separation is shown as an inset in the third column of rows (d), at binary formation, and (e), shortly before the merger.}
 \label{fig:vis2}
\end{figure*}

\subsection{Initial conditions}
\label{sec:IC}
The initial conditions of the simulation consist of two identical gas-rich dwarf galaxies on a colliding orbit. Each progenitor galaxy has a dark matter halo mass of $M_{\rm vir} = 2\times10^{10}\,\rm M_\odot$ and a rotationally supported disc with an initial gas mass of $M_{\rm gas} = 4\times10^7\,\rm M_\odot$ at metallicity $Z_{\rm gas} = 0.1\,\rm Z_\odot$, similar to the dwarf galaxy introduced in \citet{Hu_2016,Hu_2017}. These dwarf galaxies were also used in the study of IMBH growth in isolated galaxies in \citet{2025MNRAS.537..956P}. Further structural and numerical parameters are given in Table~\ref{tab:ics}. Because the simulations are non-cosmological, they do not represent a specific redshift, but the progenitor galaxy properties were originally designed to represent low-redshift dwarf galaxies.

Following \citet{2020ApJ...891....2L}, two of these galaxies are placed on a parabolic orbit with a pericentric distance of $d_{\rm peri} = 1.46\,\rm kpc$ and an initial separation of $d_{\rm init} = 5\,\rm kpc$, using the same off-plane inclinations and arguments of pericentre, $\{i_1,i_2\}=\{60\degr,60\degr\}$ and $\{\omega_1,\omega_2\}=\{30\degr,60\degr\}$. Most importantly, compared to \citet{2020ApJ...891....2L}, we reduce the dark matter softening to $\epsilon_{\rm soft} = 1 \, \rm pc$ and enhance the mass resolution to $m_{\rm res} = 400 \, \rm M_\odot$ to better resolve dynamical friction of the sinking NSCs and BHs.

In the fiducial simulation \simname{1e3-NSC}, we place a nuclear star cluster with a central IMBH at the centre of each galaxy, following, in modified form, the set-up of \citet{2025MNRAS.537..956P}. The NSCs have a total mass of $10^6 \, \rm M_\odot$ (of which $8\times 10^5 \, \rm M_\odot$ remain bound once placed in the galaxy) and a half-mass radius of $r_{1/2} = 3\, \rm pc$ ($r_{1/2} = 2.1\, \rm pc$ for bound stars), consistent with the massive end of nucleated dwarfs of comparable stellar mass \citep{Neumayer_2020,Hoyer_2023}.  The nuclear star clusters follow a \citet{Dehnen1993} profile and are realised with a uniform mass resolution of $20 \, \rm M_\odot$. We choose an initial BH mass of $10^3 \, \rm M_\odot$, which is compatible with the NSC--BH mass scaling relations as presented by \citet{Neumayer_2020, Hoyer_2024}, which show a scatter of several dex at fixed NSC mass.

As Table~\ref{tab:ics} shows, the simulations consist of five different particle types. While the force between BHs and dark matter is always gravitationally softened to account for the collisionless nature of the dark matter, the stars that form during the simulation (``new stars'', individual stars spanning the whole IMF) are always integrated with {\sc Ketju} in the vicinity of BHs, so that the BH--new-star interaction is unsoftened. During the dynamical-friction-driven NSC sinking phase, the interaction between old NSC stars and the BHs is integrated with a small softening of $0.05 \, \rm pc$ to make the simulation computationally feasible. Shortly before the NSCs coalesce, at $t_{\rm Ketju} = 460 \, \rm Myr$, {\sc Ketju} is switched on also for the NSC stars to capture the full small-scale binary black hole formation and evolution including the BH sinking inside the newly formed merged NSC and the scattering of NSC stars\footnote{We have verified that at this time the trajectories of the BHs are still almost identical with and without {\sc Ketju} and that the structural properties of the NSC are not affected.}.

With this strategy, all phases of the merger process are captured: Initially, the sinking NSC has a dynamical mass of ${\sim} 10^6 \, \rm M_\odot$, approximately $2500$ times more massive than the dark matter particles that are responsible for orbital decay in the galactic outskirts, well above the limit for resolved dynamical friction \citep[e.g.][]{Genina2024}. Later, once the NSC and the newly formed stars dominate the dynamics of the IMBH binary, the BHs have grown to $\gtrsim 3 \times 10^4 \, \rm M_\odot$, such that the hardening of the BH binary from the scattering of stars is in the well converged regime \citep[with $M_{\rm BH}/m_{\rm NSC} > 1500$, where $m_{\rm NSC}$ is the NSC particle mass, and with new stars representing individual stars rather than a sampled population;][]{Mannerkoski_2023}.

Beyond the fiducial simulation described above, we present five additional simulations as shown in Table~\ref{tab:simulations}. We test three initial conditions without a nuclear star cluster and different initial black hole masses of $10^3$, $10^4$, and $10^6 \, \rm M_\odot$ (\simname{1e3-noNSC}, \simname{1e4-noNSC}, \simname{1e6-noNSC}) to single out the effect of the NSC. To highlight the effect of the galaxy merger and the resulting starburst, we also show a simulation of a galaxy in isolation, which is identical to one of the progenitors (\simname{1e3-NSC-isolated}). Finally, we include one simulation that resets the BH mass to its initial value, shortly before the BHs form a binary (\simname{1e3-NSC-reset}).

\begin{figure*}
	\includegraphics[width=2\columnwidth]{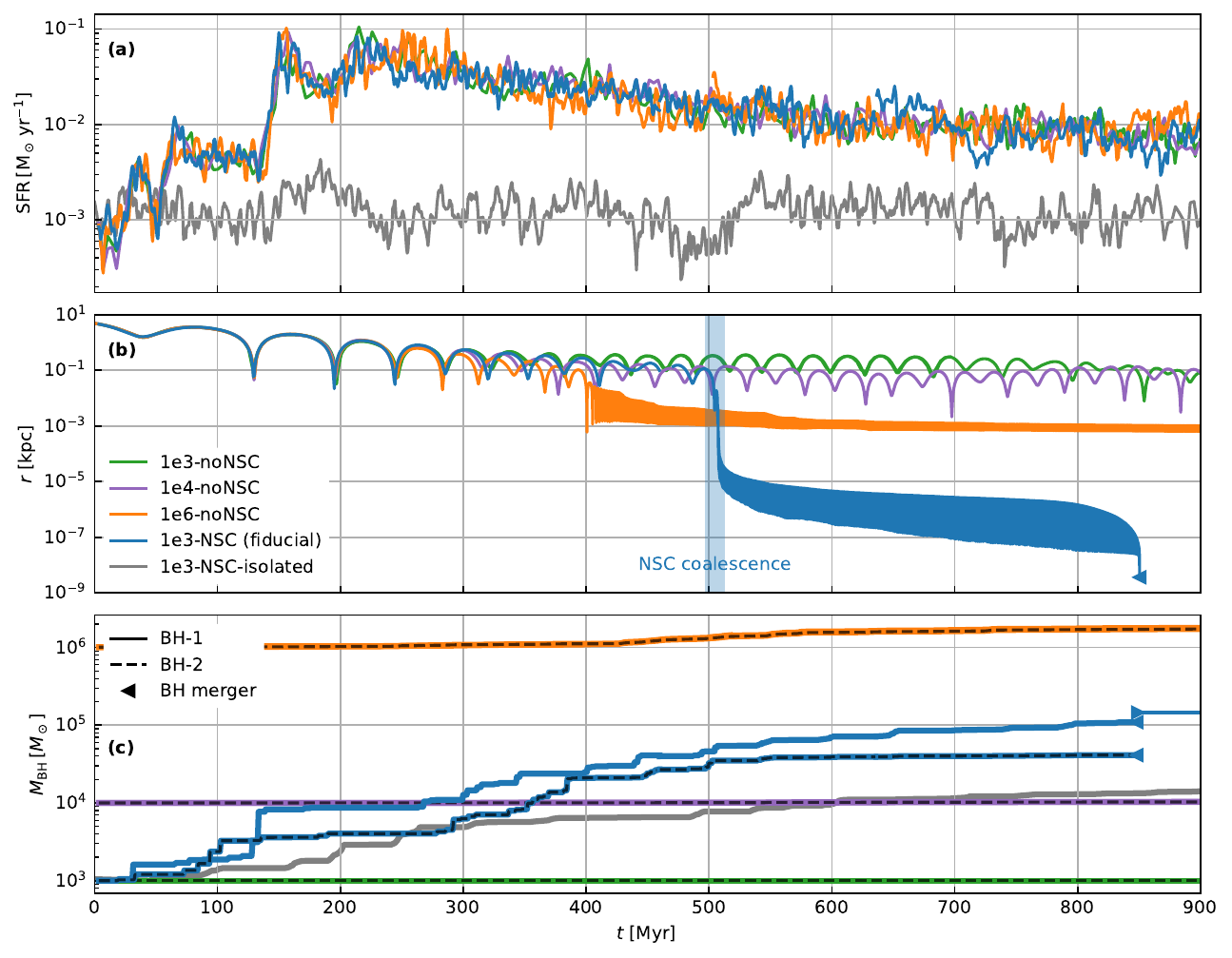}
 \caption{Evolution of the four merger simulations in comparison to the isolated simulation. The global star formation rate (SFR; panel a) shows a starburst between $150$--$300 \, \rm Myr$ that is independent of the presence of an NSC or IMBH. The isolated simulation (grey) maintains a constant star formation rate, more than one order of magnitude below the SFR in the post-starburst phase.
 Panel (b) shows the radial distance between the BHs, which also traces the distance between the NSCs (BHs always stay within $\lesssim 0.05$ pc from the NSC centre before NSC coalescence). In the fiducial simulation \simname{1e3-NSC} (blue), orbital decay due to dynamical friction leads to NSC coalescence at around $500 \, \rm Myr$ (blue shaded area). After NSC coalescence, mass segregation leads to the formation of a BH binary and stellar scattering takes over until gravitational waves cause the BHs to merge at ${\sim} 848 \, \rm Myr$. BHs without an NSC but with the mass of the fiducial NSC--BH system, $10^6\,\rm M_\odot$ (orange), initially sink faster, but then form a binary that stalls at parsec scale, with an insufficient supply of stars for stellar scattering to facilitate a fast merger. Light IMBHs without an NSC ($10^3\,\rm M_\odot$,green; $10^4\,\rm M_\odot$ purple) show very slow sinking, with their apocentric distance still at $\gtrsim 50 \, \rm pc$ at the end of the simulation. Panel (c) shows the individual masses of BH-1 and BH-2, which are indistinguishable visually in the simulations without NSC. The IMBH in the isolated galaxy (grey) grows much less than the same IMBH in the merger simulation. }
 \label{fig:sinking}
\end{figure*}

\begin{figure*}
	\includegraphics[width=2\columnwidth]{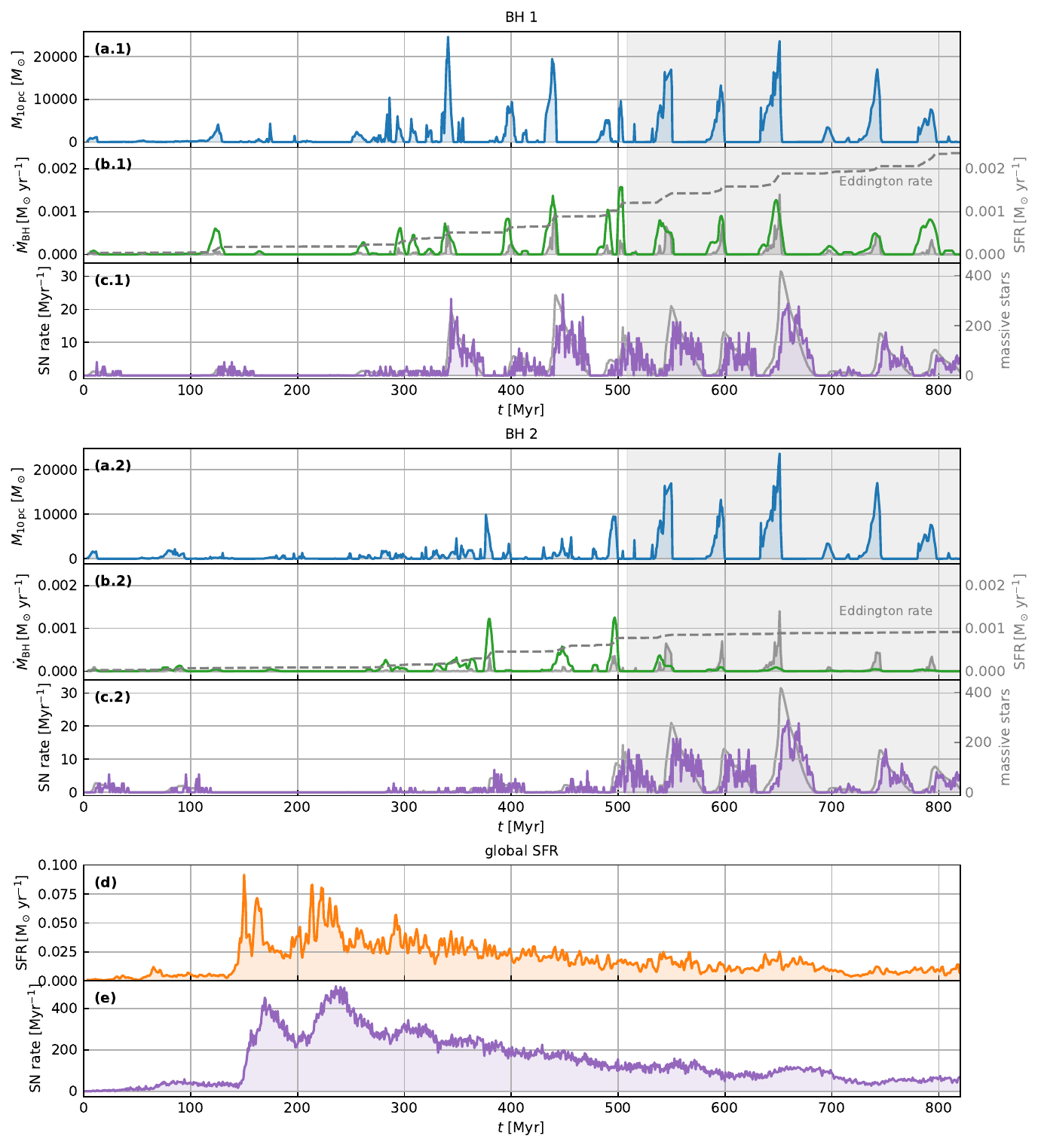}
 \caption{Evolution of the two black holes (BH-1, BH-2) in the fiducial simulation \simname{1e3-NSC} through the dwarf-galaxy merger and the nuclear starburst it triggers. Panels (a.1) and (a.2) show the gas mass within $10\,\rm pc$ of BH-1 and BH-2, respectively, tracing the nuclear fuel reservoir available for accretion and nuclear star formation.
 Panels (b.1) and (b.2) show the corresponding BH accretion rates (green) together with the Eddington-limited rate (grey dashed, rising as the BH mass grows, assuming a standard radiative efficiency of $\epsilon_{\rm r} = 0.1$), and the star formation rate within $10\,\rm pc$ (grey shaded, right axis).
 Panels (c.1) and (c.2) show the local SNe rate within $10\,\rm pc$ (purple) and the number of stars more massive than $8\,\rm M_\odot$ within $10\,\rm pc$ (grey, right axis). Outside the starburst phase, the local SNe regulate the local supply of gas by periodically ejecting gas and terminating the accretion cycles.
 The galaxy-wide star formation rate is shown in panel (d), the global SNe rate in panel (e). The grey shading marks times after NSC coalescence ($t_{\rm coal} \sim 508\,\rm Myr$).
 Gas capture by the NSCs and BHs is suppressed during the starburst and becomes episodic once the global SNe rate has decreased, after ${\sim} 340 \, \rm Myr$ (Section~\ref{sec:fbregulation}). }
 \label{fig:fbregulation}
\end{figure*}

\section{Results}
\label{sec:results}
Before presenting a comparison between the six simulations described in Table~\ref{tab:simulations}, we first provide an overview of the main evolutionary stages in the fiducial simulation, \simname{1e3-NSC}, and summarise the physical processes examined in the following sections.

\subsection{Simulation overview}

Figure~\ref{fig:vis2} shows snapshots of \simname{1e3-NSC} at different stages of the merger, illustrating the evolution and eventual coalescence of the black holes and nuclear star clusters. For each stage, we show the gas and stellar surface densities on galactic scales (columns 1 and 2) and in zoomed-in regions around one of the black holes (columns 3 and 4). The trajectories of BH-1 and BH-2 are shown in black and white, respectively. We also clarify some definitions used throughout the paper.

(a) At the first pericentre passage, the galaxies are still clearly separated and only a few new stars have formed (column 2). The pre-existing NSCs, shown with a separate colour scale, are visible as compact yellow overdensities at the galaxy centres. The stellar surface-density inset (column 4) shows that each BH remains at the centre of its host NSC.

(b) The galaxy interaction triggers a starburst, increasing the star formation rate by about two orders of magnitude relative to the isolated galaxies. Despite the high gas densities, the NSCs and BHs do not capture gas.

(c) Shortly after the galaxies have merged, the star formation rate declines and decreases further with time. We will call this phase the ``post-starburst phase''; it coincides with the phase after the coalescence of the two galaxies, but precedes the coalescence of the NSCs and of the BHs. As SN feedback subsides, the NSCs settle toward the new galactic centre and begin to capture gas, which also feeds the BHs. The NSCs have not yet merged, but their separation has decreased to ${\sim} 100\,\rm pc$.

(d) At $t_{\rm coal}\sim 508\,\rm Myr$, the NSCs have coalesced into a single bound cluster, after their pericentre separations dropped below ${\sim} 10\,\rm pc$ at ${\sim} 500\,\rm Myr$. Within the newly formed NSC (inset in column 4), the two IMBHs rapidly form a bound binary at $0.1 \, \rm pc$ scale. Three-body interactions then eject stars and shrink the binary black hole (BBH) through stellar hardening to milliparsec scale (also referred to as ``black hole scouring'' or ``BBH scattering'').

(e) Once stellar hardening has reduced the separation sufficiently, gravitational-wave emission drives the final inspiral and BH merger at $t_{\rm merge}\sim 848\,\rm Myr$. The resulting gravitational-wave recoil (``BH merger kick'') ejects the BH merger remnant from the NSC; this phase is discussed in Section~\ref{sec:merger_recoil}.

In the following sections of this paper, we will analyse the BH sinking and growth, coalescence, and evolution of the NSCs, as well as potentially observable signatures, in detail, and compare the fiducial simulation \simname{1e3-NSC} to the simulations outlined in Table~\ref{tab:simulations}.

\subsection{BH sinking and growth}
\label{subsec:growth}
In this section, we compare the fiducial simulation with the three comparison simulations. In particular, Figure~\ref{fig:sinking} shows the global star formation rate (a), the distance between the BHs (b), and the mass of the BHs (c) as a function of time for all four merger models. We also include a simulation of one of the progenitor galaxies in isolation for comparison (grey).

Irrespective of the BH mass or presence of an NSC, the simulations show a similar star formation history with a starburst between ${\sim} 150$--$300 \, \rm Myr$ induced by the interaction of the two galaxies and a decay towards the post-galaxy-merger phase, indicating that the additional potential well from the NSC or BHs does not change the global evolution of the galaxy (at least in the absence of BH feedback). In contrast, the isolated simulation \simname{1e3-NSC-isolated} has an approximately constant star formation rate 1-2 orders of magnitude lower than in the merger simulations.

In panel (b), we show the evolution of the distance between the BHs in the merger for the four different models. While initially on the same orbit, the separation of the \simname{1e3-noNSC} BHs decreases only very slowly and remains at $100$--$200\,\rm pc$. Being one order of magnitude more massive, the BHs in \simname{1e4-noNSC} sink slightly faster, but their apocentres are still at ${\sim} 50$--$100\,\rm pc$ at the end of the simulation, without forming a bound pair. This slow decay is expected: a rough estimate based on Chandrasekhar's dynamical friction formula \citep{Chandrasekhar_1943,Binney_2008}, applied to the orbits at the end of the simulation, predicts that the $10^4\,\rm M_\odot$ BHs would not become bound before ${\sim} 2\,\rm Gyr$ after the end of the simulation, and the $10^3\,\rm M_\odot$ BHs not within a Hubble time. For IMBHs of masses in this range, such slow sinking is expected and often referred to as the ``sinking'' problem \citep[e.g.][]{Bellovary_2021,Ma_2021,2024MNRAS.532.4681P}.

On the other hand, in the fiducial simulation with an NSC, \simname{1e3-NSC}, the combined NSC-BH system sinks slightly faster due to stronger dynamical friction, reaching pericentre distances between the NSCs of $\lesssim 10 \, \rm pc$ after ${\sim} 500 \, \rm Myr$. Most importantly, once the NSCs have reached these pericentre separations, the NSCs with initial half-mass radii of ${\sim} 3 \, \rm pc$ start to overlap and tidal forces cause the NSCs to merge rapidly. Inside the new cluster, dynamical friction from the NSC stars causes the two IMBHs to sink to ${\sim} 0.1 \, \rm pc$ within a few Myr, where they form a binary that merges at $t_{\rm merge}\sim 848 \, \rm Myr$. The stellar scattering phase and the gravitational-wave-driven merger will be discussed in more detail in Section~\ref{sec:Coalesence}.

In contrast, in the \simname{1e6-noNSC} simulation, where each BH is a heavy point mass, the sinking is initially faster than in the \simname{1e3-NSC} simulation and the BHs form a bound binary earlier, but the orbit stalls at pericentre distances of $\gtrsim 1 \, \rm pc$, as there is no star cluster that allows the formed binary to lose energy via stellar scattering. At this separation, the gravitational-wave inspiral time \citep{Peters_1964} exceeds the Hubble time by about ten orders of magnitude. Without a dense stellar environment, stellar hardening is also too slow to bridge this gap, as the hardening rate estimated following \citet{Quinlan_1996} implies a timescale of more than a hundred Hubble times, so that the binary would effectively never merge. Because the BHs are highly over-massive compared to the host galaxy, this simulation should be viewed as a numerical experiment to single out the effect of the NSCs in comparison to a point mass of equivalent mass. Nevertheless, it might be interesting in the context of the potentially over-massive BHs that were uncovered at high redshift, when their host galaxies were still in the dwarf-galaxy mass range \citep{Pacucci_2023,Maiolino_2024,Mezcua_2024}.

As shown in panel (c), IMBHs without an NSC do not grow efficiently. Even though the low-mass BHs (\simname{1e3-noNSC} and \simname{1e4-noNSC}) sink to within $\lesssim 100$--$200 \, \rm pc$ of the gas-rich centre, their shallow potential wells together with their residual velocity with respect to the gas in the galactic centre do not allow for the capture of gas from the turbulent ISM. A similar lack of IMBH growth has been reported in the study of isolated dwarf galaxies with IMBHs, even for IMBHs initially at rest at the galactic centre \citep{2025MNRAS.537..956P, 2025A&A...704A.177P}. Only the highly overmassive BHs in \simname{1e6-noNSC} are able to sustain gas accretion and together consume about $2.5$ per cent of the total initial gas mass of the two galaxies, but barely double their mass over the simulation time, as they already start very massive.

On the other hand, the BHs embedded into NSCs in the fiducial simulation \simname{1e3-NSC} grow episodically by $1.5$--$2$ orders of magnitude, from $1000\,\rm M_\odot$ to $1.1\times10^{5}$ (BH-1) and $4.1\times10^{4}\,\rm M_\odot$ (BH-2) prior to their merger. This is significantly faster than in the isolated galaxy (grey line), where the IMBH grows to ${\sim} 1.3 \times 10^4 \, \rm M_\odot$ in the same period of time. Interestingly, the BHs have already grown to $9000 \, \rm M_\odot$ (BH-1) and $3000 \, \rm M_\odot$ (BH-2) prior to the starburst, indicating BH activity during the approach of the galaxies on kpc scales, that is enhanced compared to the simulation of the galaxy in isolation. In contrast, during the subsequent starburst between ${\sim} 150$--$300 \, \rm Myr$, the BHs stop growing almost entirely, before growing episodically again during the post-starburst phase. This efficient growth of the IMBHs prior to their pairing as a binary also affects the BH coalescence inside the NSC, as we will discuss in Section~\ref{sec:Coalesence}. The final mass of the merger remnant is more than two orders of magnitude larger than the seed mass, and more than one dex more massive than in the isolated simulation.

\begin{figure*}
	\includegraphics[width=2\columnwidth]{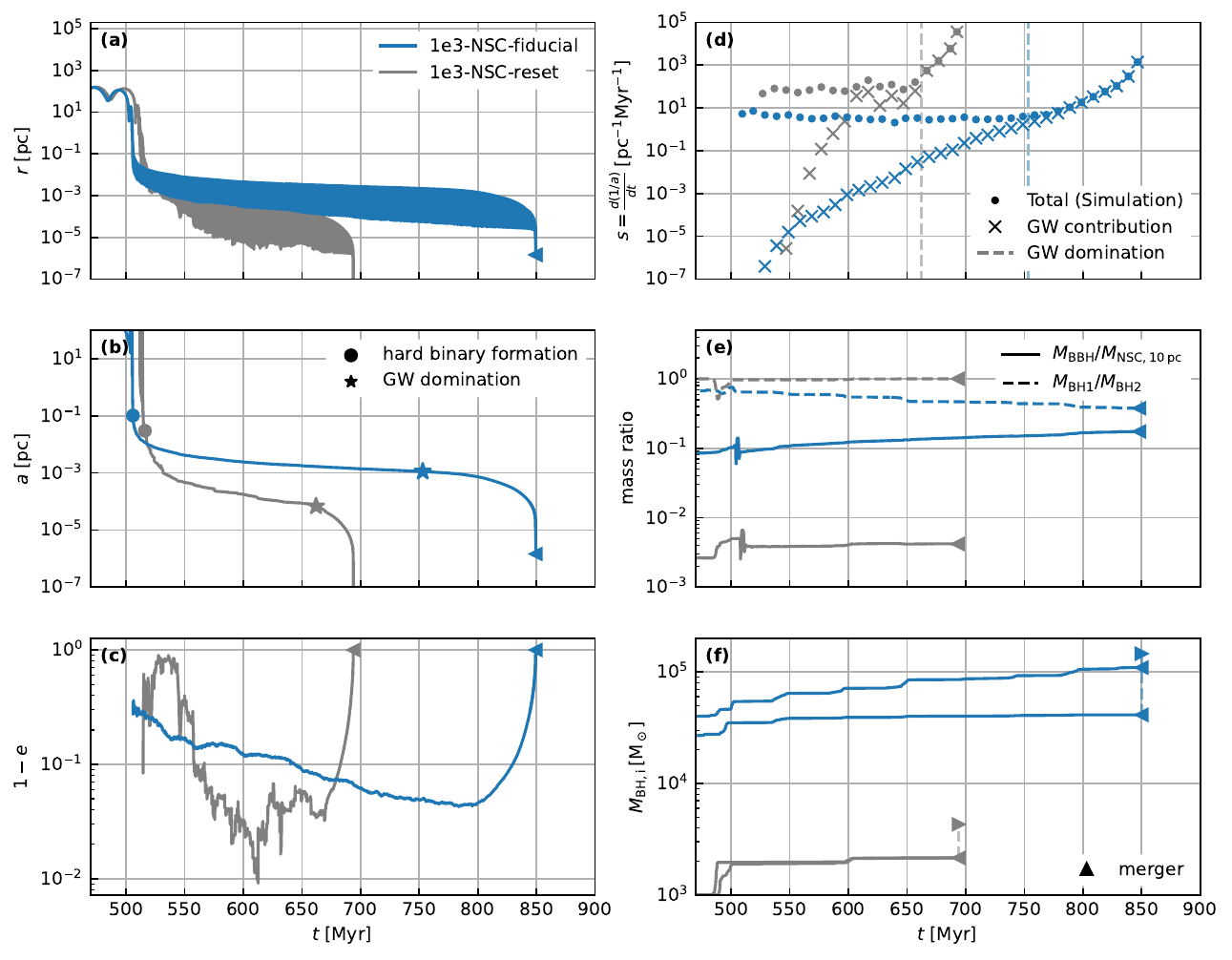}
 \caption{Dynamical evolution of the BH pair in the fiducial simulation (\simname{1e3-NSC}, blue): The figure shows (a) pairwise separation $r$; (b) semi-major axis $a$ once the BHs form a bound binary, with the triangle marking the merger; (c) eccentricity, shown as $1-e$; and (f) mass of each BH, which grows via gas accretion and through the merger at final time (the right-pointing triangle marker indicates the merger remnant mass).
 Panel (d) shows the total hardening rate $s_{\rm total}$ as measured from the simulation, and its gravitational-wave contribution $s_{\rm GW}$ separately. The time at which gravitational waves take over is shown as a dashed vertical line (and a star marker in panel b).
 The BBH-to-NSC mass ratio and the BH mass ratio $q=M_2/M_1\leq1$ are shown in panel (e).
 For comparison, we add the numerical experiment \simname{1e3-NSC-reset} (grey) in which the BH mass is instead reset to $10^3\,\rm M_\odot$ at $t\sim460\,\rm Myr$ (Section~\ref{sec:Coalesence}).}
\label{fig:coalescence}
\end{figure*}

\subsection{Nuclear gas supply and stellar feedback regulation}
\label{sec:fbregulation}

To understand the episodic IMBH growth of the fiducial simulation \simname{1e3-NSC} (Figure~\ref{fig:sinking}), we analyse the gas content in the vicinity of the BH and the correlation with stellar feedback in Figure~\ref{fig:fbregulation}. For both BHs (1 and 2), we show the supply of gas within $10\,\rm pc$ as a tracer for the available material for nuclear star formation and BH accretion (a). The BH accretion rate as computed from our sink particle prescription\footnote{With the sink particle prescription accreting discrete particles without feedback, the accretion rate itself is not dynamically important during the simulation and is computed by smoothing the BH mass as a function of time with a running mean over $2\,\rm Myr$ to reduce discreteness noise from individual accretion events.} is shown in panel (b), together with the star formation rate within $10\,\rm pc$ (grey, right axis), and the SN rate within $10\,\rm pc$ (c), together with the number of stars more massive than $8\,\rm M_\odot$ within $10\,\rm pc$ (grey, right axis), i.e. the SN progenitors present at any time. In addition, we show the global SFR and the global SNe rate in panels (d) and (e).

While there are short episodes of accretion, nuclear star formation and SN feedback in the pre-merger phase ($t < 150 \, \rm Myr$), there is essentially no BH growth in the starburst phase ($150$--$300 \, \rm Myr$). During this phase, even though the BHs are passing through the centre of the starburst with high gas densities regularly and spend time in the dense regions (see also the inset in row b of the overview Figure~\ref{fig:vis2}), the BHs and their associated NSCs do not capture and hold on to gas (panels a.1 and a.2).

While it is difficult to disentangle the effects quantitatively, we attribute this suppression to the high global energy injection rate from SNe (panel e), which produces a highly turbulent ISM, and to the residual streaming velocity of the NSCs with respect to the surrounding gas.

In particular, we find that during this phase, the NSCs are still moving together with their dark matter environment ($|\Delta v_{\rm NSC-dm}| \lesssim 10 \, \rm km\,s^{-1}$), which they inherited from the progenitor galaxies. The gas, in contrast, has decoupled from its former host galaxy and from the NSCs. During the starburst, the velocity of the NSC with respect to the gas in its $100 \, \rm pc$ environment, i.e. its streaming velocity, varies between $40$--$200 \, \rm km\,s^{-1}$. The gas itself, measured inside $r < 100 \, \rm pc$ around the BHs, is highly turbulent with a velocity dispersion of the order of ${\sim}20$--$100 \rm \, km\,s^{-1}$ during the starburst.

These velocities are comparable to the escape velocity from the NSC at its half-mass radius (${\sim} 50 \, \rm km\,s^{-1}$) and several times the escape velocity at $10$ and $100 \, \rm pc$ ($26$ and $8 \, \rm km\,s^{-1}$, respectively), where most gas actually resides. Hence, when the NSCs plunge through the centre of the galaxy, where the gas density is highest, their velocity is too high to capture gas from the ISM (even if it is in the cold ISM phase with sound speed $\lesssim 1 \, \rm km\,s^{-1}$).

In contrast, as soon as the starburst decays and the NSCs settle in the galactic centre (at ${\sim} 300$--$340\,\rm Myr$), the NSCs episodically capture gas from the ISM. During this phase, the typical streaming velocity of the gas with respect to the NSC has settled at $\lesssim  20 \, \rm km\,s^{-1}$ (velocity dispersion $\sigma \lesssim 30 \, \rm km\,s^{-1}$), qualitatively explaining why gas capture is possible now.

The captured gas efficiently feeds the black holes on timescales of ${\sim} 10$--$30 \, \rm Myr$ and leads to star formation inside the NSC. The SNe from stars that formed inside the clusters take over the regulation of the BH growth, by expelling gas from the centre after each cycle of BH accretion, as soon as the first massive stars die. This is similar to the isolated simulation in \citet{2025MNRAS.537..956P}, where the onset of SNe (here the purple curve in panel c) periodically terminates nuclear star formation and BH accretion cycles by expelling gas from the NSC. In Figure~\ref{fig:fbregulation}, these cycles are visible as a correlation between the gas supply and BH accretion rate (panels a and b), which both decrease to zero as soon as the nuclear SNe rate increases shortly after the onset of nuclear star formation. The nuclear star formation rate (grey in panels b.1 and b.2) peaks together with the accretion rate, the number of massive stars within $10\,\rm pc$ (grey in panels c.1 and c.2) rises within a few Myr, and the first SNe of these stars terminate the cycle.

During periods of efficient BH growth, the accretion rate onto the sink particles can generally reach super-Eddington rates for a few Myr (no Eddington limit is imposed in the sink-particle prescription). The BH masses increase over time due to gas accretion (see panel c in Figure~\ref{fig:sinking}), such that their spheres of influence grow, gas becomes bound to the BHs more easily, and they accrete a larger fraction of the gas in their vicinity. However, the gas supply enclosed within $10\,\rm pc$ of the BHs is mostly independent of the BH mass, as the NSC ($M_{\rm NSC} \sim 10^6 \, \rm M_\odot$) is more than one order of magnitude more massive than the BHs. In the final phase, when the BHs form a binary, both BHs have the same $10\,\rm pc$ gas supply (grey shaded region in Figure~\ref{fig:fbregulation}, which marks the time after the coalescence of the NSCs). In this stage, as a consequence of the gravitational boundedness criterion to each individual sink particle in the accretion prescription, the more massive BH accretes more gas.

Hence, in the simulation, the time period for most efficient BH growth is not the starburst with high gas densities itself, but the post-starburst phase where the BHs and their NSCs have settled in the galactic centre, the global SNe rate has decreased and the nuclear gas supply becomes regulated by local star formation and the associated SN feedback.

\begin{figure}
	\includegraphics[width=1\columnwidth]{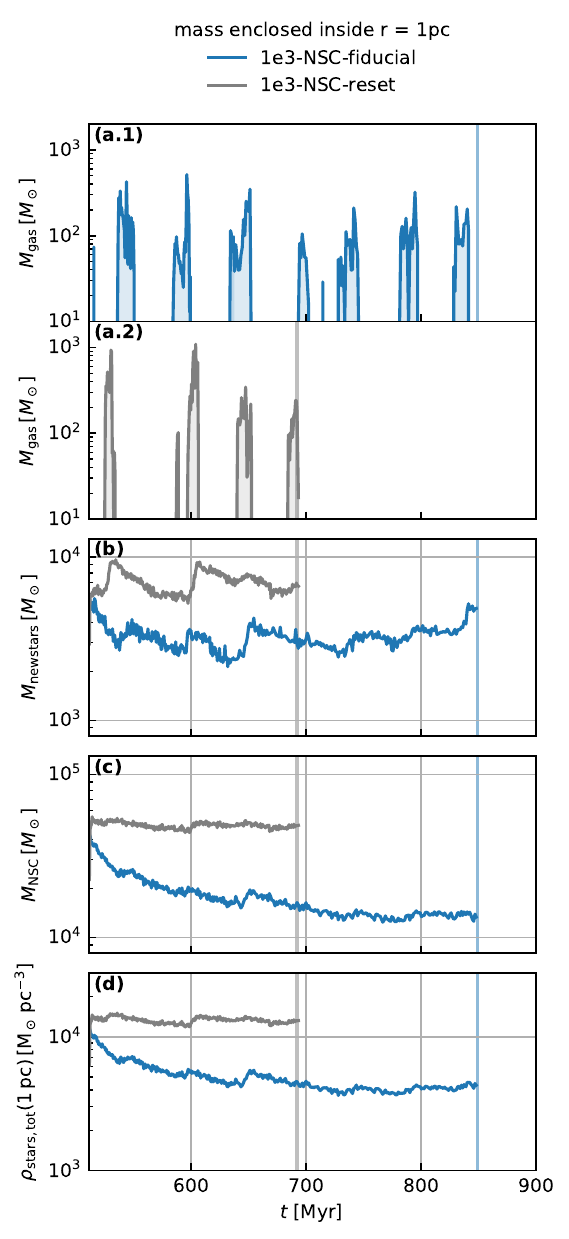}
 \caption{Nuclear stellar and gas content within $r=1\,\rm pc$ of the BH binary in the fiducial simulation \simname{1e3-NSC} (blue), with the mass-reset numerical experiment of Figure~\ref{fig:coalescence}, \simname{1e3-NSC-reset}, for comparison (grey): Panels (a.1) and (a.2) show the gas mass for both simulations, each showing repeated episodes of local gas capture and depletion (see the SNe regulation in Figure~\ref{fig:fbregulation}); (b) shows mass in newly-formed stars that depletes due to stellar scattering and stellar evolution mass-losses; (c) mass in old, pre-existing NSC stars. The NSC density drops due to BH binary scattering in the fiducial run, but is left essentially unperturbed by the much lighter BBH in \simname{1e3-NSC-reset}; (d) resulting mean total stellar mass density within $1\,\rm pc$, combining BBH scattering, star formation, and stellar-evolution mass loss. The vertical blue/grey lines mark the fiducial/reset binary's coalescence time.}
 \label{fig:available_supply}
\end{figure}

\subsection{BH coalescence and stellar scattering}
\label{sec:Coalesence}

In this section, we continue to analyse the coalescence of the NSCs and BHs in the fiducial simulation, \simname{1e3-NSC}, and the final stage of the binary black hole (BBH) merger. To explore the effect of the BH mass on BH coalescence, we also conduct a numerical experiment, \simname{1e3-NSC-reset}, where we reset the masses of BH-1 and BH-2 back to $1000 \, \rm M_\odot$ each at $t = 460 \, \rm Myr$ by hand (from $4 \times 10^4$ and $2.8 \times 10^4 \, \rm M_\odot$, respectively), shortly before the NSCs merge.\footnote{At this time, {\sc Ketju} integration is also switched on for the old NSC stars (Section~\ref{sec:IC}).}

In Figure~\ref{fig:coalescence} we show the pairwise separation $r$ (panel a), the semi-major axis $a$ of the binary once the BHs are bound to each other (panel b), and its eccentricity ($1-e$; panel c).

In the fiducial simulation \simname{1e3-NSC} (blue), the binary becomes bound shortly after the NSCs have merged. Rapid sinking (mass segregation) inside the NSC leads to the formation of a hard binary, as defined by $a < a_{\rm h}=[q/(4(1+q)^2)]\,r_{\rm infl}$, where $q=M_2/M_1\leq1$ and the binary sphere-of-influence radius $r_{\rm infl}$ encloses a stellar mass equal to twice the total binary mass.

After the formation of a hard binary, stellar scattering shrinks the BBH semi-major axis with an approximately constant hardening rate $s_{\rm total} \equiv \frac{{\rm d}}{{\rm d}t}\left(\frac{1}{a}\right)$ (panel d, measured from the derivative of the semi-major axis in panel b). Starting from an initial orbit with a moderate eccentricity of ${\sim} 0.7$, the eccentricity of the binary grows to $e = 0.96$ during the BBH scattering phase. At the same time, BH growth via gas accretion drives up the combined BBH mass from $M_{\rm BBH} = M_1 + M_2 \sim 8 \times 10^4 \, \rm M_\odot$ to $1.5 \times 10^5 \, \rm M_\odot$ since the coalescence of the NSCs (panel f). As a result, the mass ratio $M_{\rm BBH}/M_{\rm NSC, 10\,pc}$ (with $M_{\rm NSC,\,10\,pc}$ measured within $10\,\rm pc$) increases from ${\sim} 10$ to ${\sim} 17$ per cent, making the BBH progressively more dynamically important inside the NSC. As the binary grows, the orbital energy that stellar scattering has to extract before gravitational-wave emission takes over increases in proportion to $M_{\rm BBH}$, and so does the stellar mass that the binary must eject \citep{Merritt_2006}.

At ${\sim} 750 \, \rm Myr$, the post-Newtonian terms responsible for gravitational-wave emission start to dominate the hardening rate (i.e. the post-Newtonian hardening rate, which is self-consistently followed in the simulation, contributes more than $50$ per cent to the total hardening rate). This causes the total hardening rate in panel (d) (Figure~\ref{fig:coalescence}) to go up, while the binary circularises and finally reaches the merger criterion of $100 \, R_{\rm S}$ at $t_{\rm merge}\sim 848 \, \rm Myr$.

The mass reset in \simname{1e3-NSC-reset} keeps the BBH artificially light at the onset of the BH coalescence phase. As Figure~\ref{fig:coalescence} shows, this choice translates into more than an order-of-magnitude difference in how massive the BBH is relative to its host NSC, $M_{\rm BBH}/M_{\rm NSC, 10\,pc}\sim 10$--$17$ per cent in the fiducial run versus ${\sim} 0.4$--$0.5$ per cent in the reset run, making this comparison a controlled test of how coalescence depends on the BBH-to-NSC mass ratio, rather than on the BH growth history prior to the NSC merger.

Because the lighter BHs experience weaker dynamical friction and mass segregation, the BBH in the comparison run (\simname{1e3-NSC-reset}, grey) reaches the hard-binary stage a few Myr later than in the fiducial simulation. However, because the BBH remains light relative to the NSC throughout, stellar scattering does not deplete the cusp of the NSC fast and $a$ continues to shrink smoothly with a hardening rate approximately one order of magnitude higher than in the fiducial case. With an eccentricity of up to ${\sim} 0.99$, the BBH enters the GW dominated regime at ${\sim} 660 \, \rm Myr$ and crosses the merger threshold at $t\sim690\,\rm Myr$, some $160\,\rm Myr$ before the fiducial run.

We note that at approximately constant BH masses of ${\sim} 2000 \, \rm M_\odot$ (panel f) during the longest part of the BH coalescence, the BH masses are only about one order of magnitude higher than the masses of most massive stars sampled from the \citet{2001MNRAS.322..231K} IMF, so that the stellar mass function and the internal structure of the NSC become increasingly important for the BH coalescence process. In Figure~\ref{fig:coalescence}, this shows up as small jumps in the semi-major axis and eccentricity due to the scattering of individual massive stars and nuclear star cluster stars at uniform $20 \, \rm M_\odot$ resolution. We also note that the instantaneous decrease of the BH masses causes the inner part of the NSC to expand, reducing the central density of the NSC compared with its state before the reset.

\begin{figure}
	\includegraphics[width=1\columnwidth]{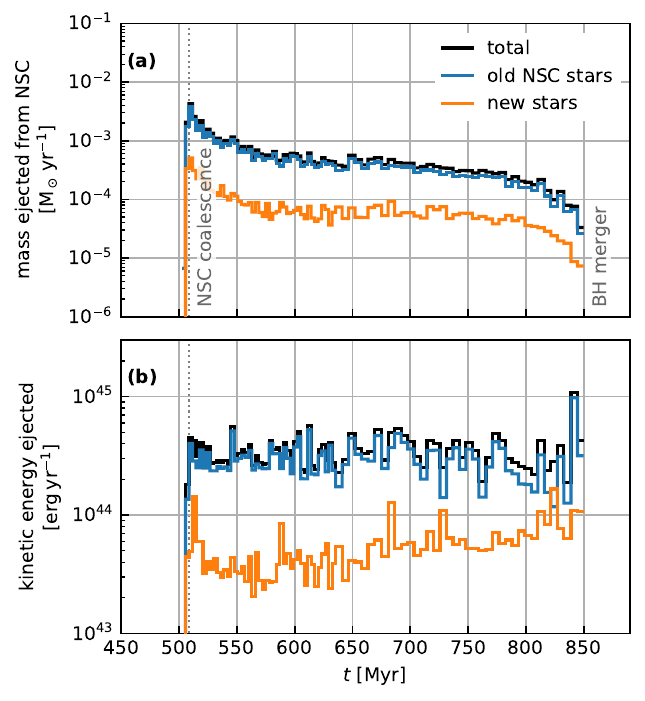}
 \caption{Mass ejection rate from the nuclear star cluster in the fiducial simulation \simname{1e3-NSC} as a function of time. To qualify as an ejection, a star must be bound to the NSC at time $t$ and unbound at $t + \Delta t$, with $\Delta t = 7 \, \rm Myr$, at a distance of more than $200 \, \rm pc$ from the NSC. The mass ejection rate (panel a) peaks directly after the formation of the BH binary and then decays towards the BH merger, while the rate of kinetic energy carried away by the ejected stars (panel b) remains approximately constant. }
 \label{fig:unbinding}
\end{figure}

To understand how the binaries interact with their environment and what drives the BH coalescence in the two simulations, we show the nuclear stellar and gas supply within $1\,\rm pc$ of the BHs in Figure~\ref{fig:available_supply}. In the fiducial run, the total stellar density within this radius drops by more than a factor of 2 over the simulation (panel d), as the BBH ejects nuclear stars faster than they can be replenished by star formation. Old, pre-existing NSC stars dominate this reservoir throughout (panel c), over the stars that form in situ during the simulation (panel b). Gas that is episodically captured within $1\,\rm pc$ (panel a.1) fuels bursts of star formation roughly every ${\sim} 50\,\rm Myr$, but the resulting young stellar mass is quickly lost again -- to stellar evolution (winds and SNe) and to dynamical ejections by the binary -- producing the saw-tooth evolution of the mass of nuclear new stars and the total stellar density (panels b and d). However, at any given time, the mass of new stars remains subdominant compared to the pre-existing NSC stars and the hardening rate does not measurably increase after star formation cycles in Figure~\ref{fig:coalescence}, panel d.

\begin{figure}
	\includegraphics[width=1\columnwidth]{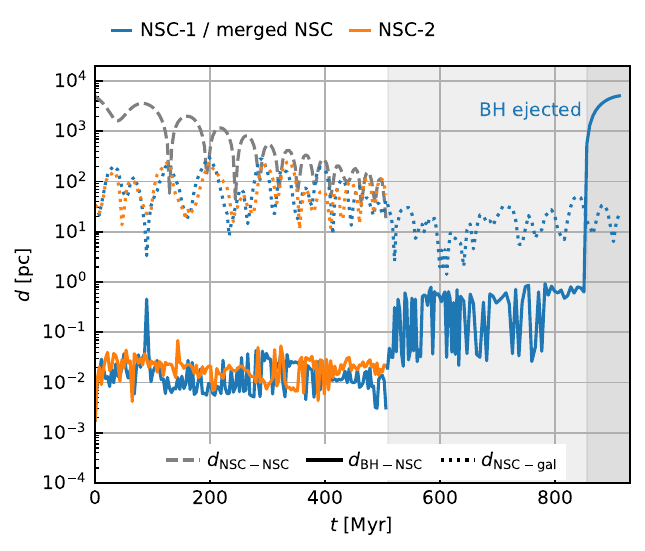}
 \caption{Evolution of the distance of the NSCs to each other, the NSCs to their respective host galaxies (centre of the dark matter initially associated with each host galaxy), and the distance between NSCs and their BHs in simulation \simname{1e3-NSC}. The GW recoil leads to the ejection of the merger remnant from the nuclear star cluster, but it is retained inside the galaxy. The light and dark grey shaded regions mark the time after the coalescence of the NSCs and after the BH merger, respectively. }
 \label{fig:recoil}
\end{figure}

In \simname{1e3-NSC-reset}, by contrast, the stellar density within $1\,\rm pc$ stays approximately constant, consistent with a BBH too light to meaningfully deplete its surrounding NSC.

\begin{figure*}
	\includegraphics[width=2.0\columnwidth]{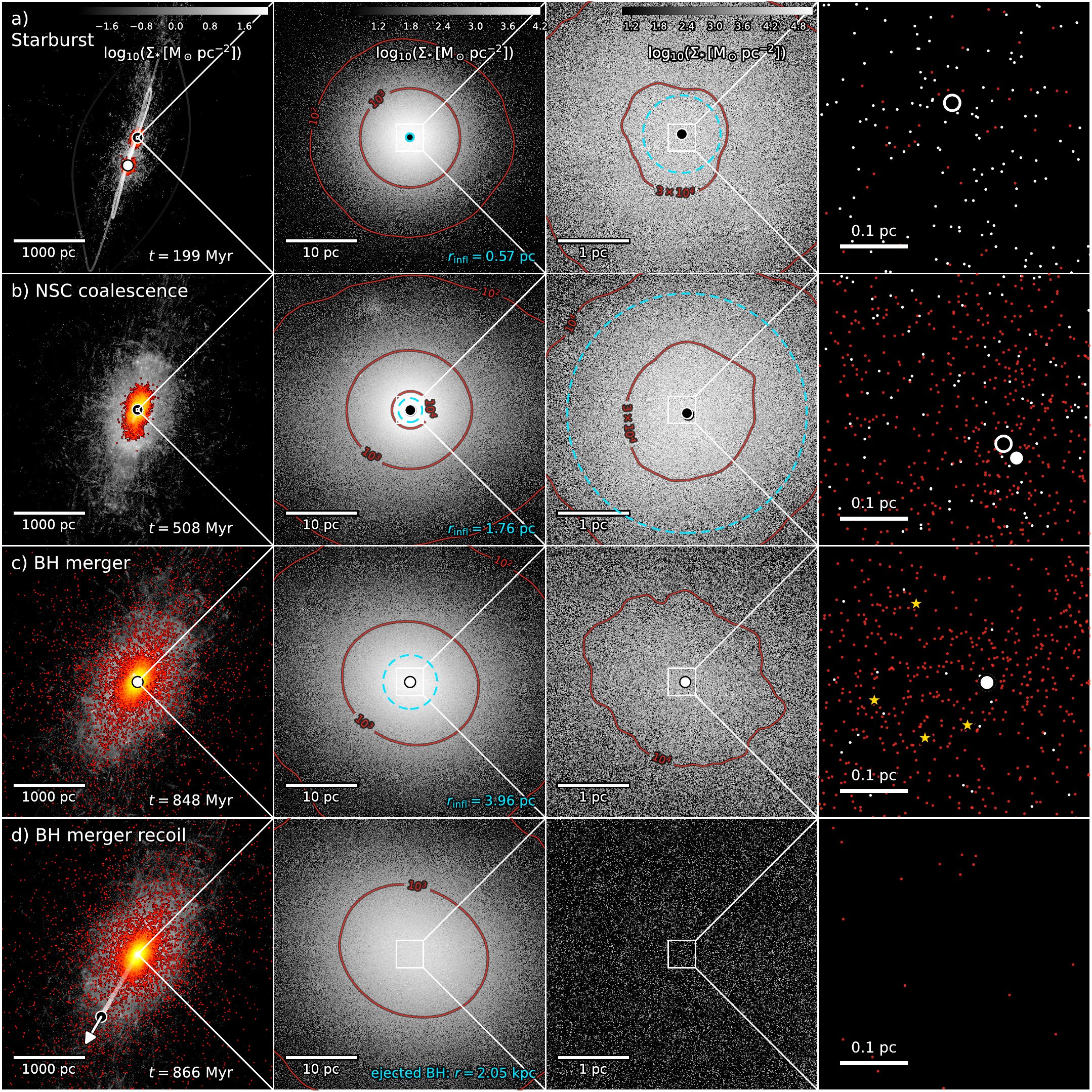}
 \caption{Evolution of the NSC during the galaxy merger, (a) before NSC coalescence, (b) after NSC coalescence but before hard binary formation, (c) during the BBH stellar scattering phase shortly before the merger, and (d) after the merger with BH recoil.
 The first column shows the surface density of new stars over-plotted by the (old) NSC surface density (as in Figure~\ref{fig:vis2}). Columns two and three show the total surface density (old NSC + new stars), with contours of equal surface density in red. Column four shows individual stars in the $0.4 \, \rm pc$ around the BH with NSC stars plotted in white, new stars in red, and massive new stars ($m > 8 \, \rm M_\odot$) in yellow. The locations of the BHs are shown as white and black dots, and the dashed cyan circles mark the sphere of influence $r_{\rm infl}$ of the BHs.
 The NSC merger (b, $t = 508 \, \rm Myr$) adds mass to the outskirts of the NSC, such that the $10^2 \, \rm M_\odot\,pc^{-2}$ contour moves to larger radii. During the scouring phase ($508$--$848 \, \rm Myr$), the central density decreases and the $\Sigma = 10^4 \, \rm M_\odot\,pc^{-2}$ contour moves inwards. Ejected stars from the nuclear star cluster are visible as red dots scattered throughout the galaxy in the overview panels in rows (c) and (d). The BH merger recoil unbinds the NSC core on the scale of its sphere of influence ($r_{\rm infl} = 3.96 \, \rm pc$ before the ejection) and leaves behind a diffuse stellar core with surface density $\ll 10^4 \, \rm M_\odot\,pc^{-2}$.   }
 \label{fig:ejected_nsc_stars}
\end{figure*}

\subsubsection{Impact of new stars on BH binary hardening}

While Figure~\ref{fig:available_supply} shows the supply of stars available for scattering by the BBH, it does not quantify the rate at which stars are actually scattered from the pre-existing NSC population and from the newly formed stars. To quantify the contribution of both populations to the BBH hardening, we determine the rate of BBH scattering events as a function of time in Figure~\ref{fig:unbinding} for the fiducial simulation \simname{1e3-NSC}.

To identify scattering events from the simulation data, we use {\sc subfind} \citep{Springel2001} with the friends-of-friends algorithm \citep[][with $1 \, \rm pc$ linking length]{Davis1985} for identifying the bound NSC. If a star is bound at time $t$ and unbound $7 \, \rm Myr$ later with a distance to the NSC of $>200 \, \rm pc$ (such that $v \gtrsim \rm 30 \, km\,s^{-1}$), we count the star as scattered. While this method cannot check explicitly whether the star had a close encounter with the BBH, there are no other processes in the simulation that lead to such fast ejections, as confirmed by the observation that no stars escaped prior to the BBH formation. We also checked that reasonable variations of the distance and time thresholds do not affect the results. In addition, we show the energy carried by ejected stars in panel (b) of Figure~\ref{fig:unbinding}. Because stars have already left the nuclear potential well at the time of the energy measurement ($>200 \, \rm pc$), we can only measure the remaining energy and this analysis provides a lower limit on the energy loss of the BBH.

The scattering rate is highest shortly after the binary forms and drops over time, as the cross-section for strong three-body encounters shrinks in proportion to the semi-major axis of the hardening binary \citep{Quinlan_1996,Sesana_2007}. However, the energy carried by these scattering events remains approximately constant over time, because the ejection velocity is of the order of the orbital velocity of the BH binary, which increases over time \citep{Quinlan_1996,Merritt2005}. At all times, new stars account for roughly $20$ per cent of the scattered mass (and carry a similar fraction of energy), indicating that ongoing star formation provides a relevant, but overall subdominant contribution to the hardening compared to the pre-existing, dense, NSC population. By analysing the age distribution of scattered stars, we do not find evidence for the preferential scattering of young stars. This likely explains why the hardening rate remains approximately constant in Figure~\ref{fig:coalescence}, independent of recent star formation episodes.

Taken together, the fiducial simulation shows that IMBH coalescence driven by stellar scattering and gravitational waves is possible within ${\sim} 340 \, \rm Myr$ after the NSCs merge, with the old NSC population dominating the scattering rate.

\subsection{Black hole merger recoil}
\label{sec:merger_recoil}
When BHs merge, general relativity predicts a gravitational-wave recoil kick that depends on the mass ratio and the spins of the BHs \citep{Lousto2010}. For simplicity, the BH spins are set to zero initially and do not evolve in the simulation, while the masses evolve over time via gas accretion. In \simname{1e3-NSC}, the masses at the time when the merger criterion is fulfilled are $M_1 = 1.086 \times 10^5$ and $M_2 = 4.119 \times 10^4 \, \rm M_\odot$. For this mass ratio of $q = 0.38$, the simulation assigns a BH recoil kick of $174.9 \, \rm km\,s^{-1}$ to the merger remnant, using the fitting formulae from \citet{2015PhRvD..92b4022Z}, which are calibrated on numerical-relativity simulations of unequal-mass, precessing binaries and give the remnant mass, spin and recoil velocity as functions of the mass ratio and the spin vectors of the progenitors.

To analyse the effect of the merger recoil on the BH trajectory, we show the evolution of the distance between the BHs and their host NSC in the fiducial simulation \simname{1e3-NSC} in Figure~\ref{fig:recoil}. Before NSC coalescence, BHs stay within $\lesssim 0.05 \, \rm pc$ of the centre of their host NSC (all centres in this paper are determined with the shrinking-sphere algorithm of \citealt{Power_2003}). When the NSCs merge, the central density profile of the NSC becomes shallower and the BH binary starts to wander up to $\lesssim 1 \, \rm pc$. Finally, the BH recoil kick ejects the remnant from the NSC. For reference, at the time of the merger, the escape velocity from the centre of the NSC to $50 \, \rm pc$ is ${\sim}50 \, \rm km\,s^{-1}$, while the escape velocity from the host galaxy is ${\sim}263 \, \rm km\,s^{-1}$. Hence, the merger remnant in simulation \simname{1e3-NSC} with a recoil kick of $174.9 \, \rm km\,s^{-1}$ is ejected from the nuclear star cluster, but remains bound to the host galaxy \citep{Blecha_2008,Gualandris_2008,Holley-Bockelmann_2008,Rawlings_2025} and will not sink back to the galactic centre within a Hubble time, since the dynamical-friction time of the remnant on its ${\sim} 5\,\rm kpc$ orbit exceeds $10\,\rm Gyr$. No stars remain bound to the ejected merger remnant. The ejection also removes the BH from the core of the NSC, which especially affects the former sphere of influence of the BBH, where it accounts for one third of the enclosed mass by definition of $r_{\rm infl}$ ($M_\star(<r_{\rm infl}) = 2M_{\rm BH}$). By the virial theorem, the instantaneous loss of mass leaves the remaining stars with an excess of kinetic energy, and they settle at a larger radius \citep[cf.][]{Hills_1980,Naab_2009}, in line with the expansion of the cluster radius discussed in Sections~\ref{sec:NSCevo} and \ref{sec:mock_photometry}.

In contrast, the BHs in \simname{1e3-NSC-reset} have a mass ratio of $q \sim 0.999$ with a recoil kick velocity of $0.273 \, \rm km\,s^{-1}$, such that the merger remnant stays inside the NSC.

\begin{figure}
	\includegraphics[width=1.0\columnwidth]{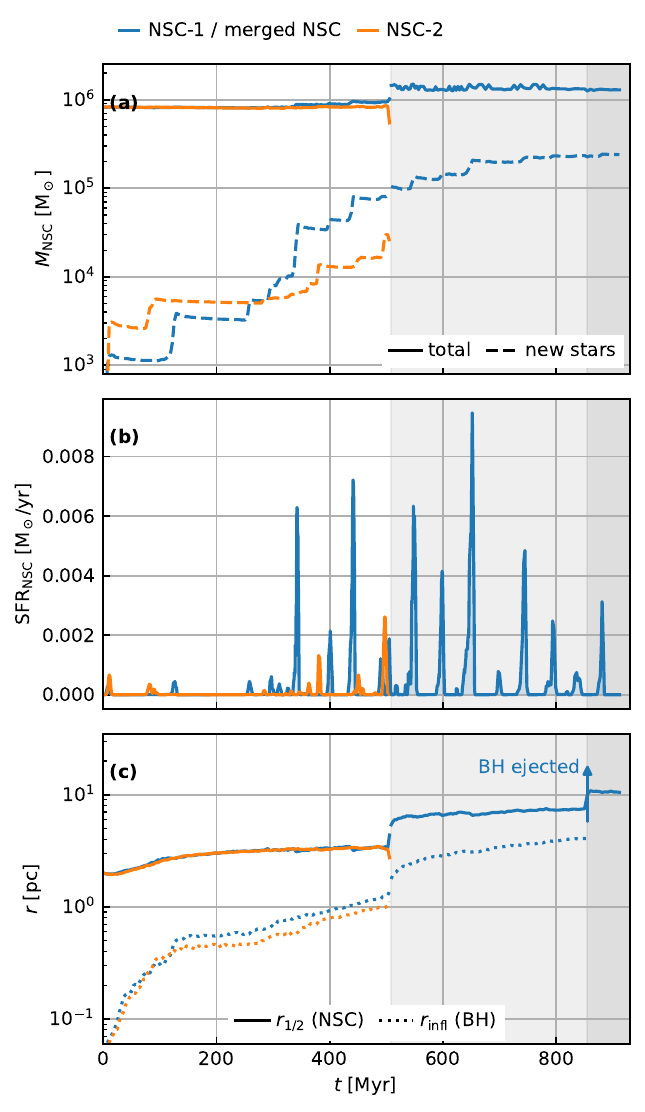}
 \caption{Evolution of the NSC. Panel (a) shows the evolution of the bound mass of the two NSCs and the contribution from new stars. New stars form episodically when gas becomes bound to the NSC (b). The size evolution of the NSC (half-mass radius $r_{1/2}$ and the BH sphere of influence $r_{\rm infl}$) is shown in (c). The distances between the NSCs, their host galaxies and their BHs are shown in Figure~\ref{fig:recoil}.}
 \label{fig:NSC_evolution}
\end{figure}

\begin{figure}
	\includegraphics[width=1\columnwidth]{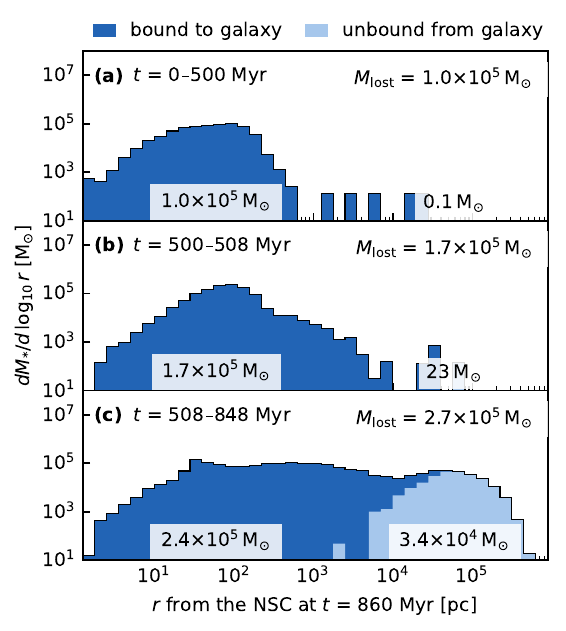}
 \caption{Histogram of the radial distribution of stellar mass that was last bound to one of the NSCs before NSC coalescence (a), during the NSC coalescence (b), and during the BH binary scattering phase (c). The plot hence indicates where stellar mass that was bound to the NSC at any time is re-distributed in the galaxy. $M_{\rm lost}$ is the mass that became unbound from either of the two NSCs during the respective period of time. During the NSC merger and the onset of BBH scattering, mass is distributed to kpc scales (via tidal stripping), but remains bound to the galaxy. During the BBH scattering phase, about 20 per cent of the cluster mass is lost, and ${\sim} 2.5$ per cent of the NSC mass is ejected and becomes unbound from the galaxy.}
 \label{fig:histogram}
\end{figure}

\begin{figure*}
	\includegraphics[width=2\columnwidth]{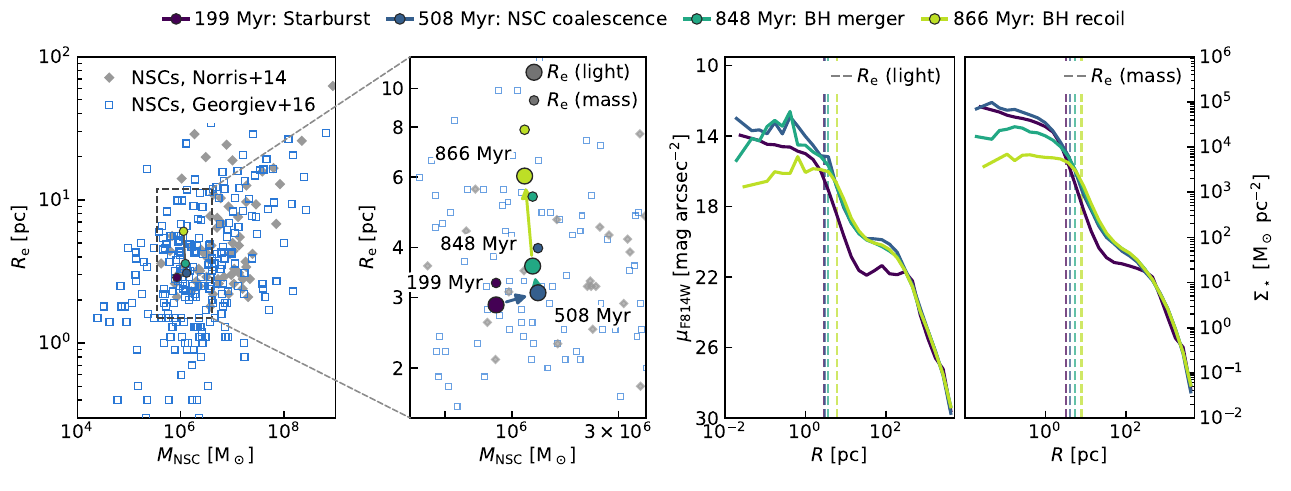}

 \caption{Structure of the simulated NSC in \simname{1e3-NSC} during the starburst phase ($199\,\rm Myr$), the NSC coalescence ($508\,\rm Myr$), the BH merger ($848\,\rm Myr$) and $18\,\rm Myr$ after the recoil kick ($866\,\rm Myr$). The left panel shows the mass--size relation for observed NSCs \citep[][NSCs only]{Norris_2014,Georgiev_2016} and the simulated NSC. In the second panel, large and small circles are the effective radii of the NSC component of two-component (NSC + host) S\'ersic fits to the projected F814W surface brightness and stellar mass profiles, respectively, and $M_{\rm NSC}$ is the integrated mass of the NSC component of the mass fit. The third panel shows the F814W surface brightness profiles and the right panel the stellar mass surface density profiles, both projected along $z$. Dashed vertical lines mark the respective NSC effective radii. }
\label{fig:nsc_obs}
\end{figure*}

\subsection{Evolution of the NSC}
\label{sec:NSCevo}

Finally, we analyse how the NSCs in the fiducial simulation \simname{1e3-NSC} evolve during the merger process. To find implications for NSC growth and potentially observable signatures of the processes described in the previous sections, we examine how the NSC mass and size evolve, and trace where stars that become unbound from the NSC during different phases of the merger process are redistributed inside the galaxy.

An overview of the NSC evolution is given in Figure~\ref{fig:ejected_nsc_stars}, which shows the stellar surface density on different spatial scales through different phases of the galaxy merger. The surface density maps show a clear increase of the NSC size during the NSC coalescence (panel a vs b, $10^2 \, \rm M_\odot\,pc^{-2}$ contour), the ejection of NSC stars from the nucleus during the BH binary phase (panel c, red dots), and the ejection of the BBH merger remnant (panel d, white arrow) that decreases the central density in the NSC (panel d, third column).

\subsubsection{NSC mass and size evolution}
The quantitative evolution of the NSC is shown in Figure~\ref{fig:NSC_evolution}. Panel (a) shows the bound mass of the NSC. Initially, the bound mass is approximately constant with $M_{\rm NSC} = 8.2 \times 10^5\,\rm M_\odot$. When the NSCs merge, the new cluster has a combined bound mass of $M_{\rm NSC} = 1.37 \times 10^6 \, \rm M_\odot$, indicating that about $2.7 \times 10^5 \, \rm M_\odot$ were stripped before and during the merger of the NSCs (see also panel a and b in Figure~\ref{fig:histogram}).

While the mass of the NSC is dominated by old NSC stars at all times, there is a contribution to the mass growth from in-situ star formation inside the NSC (panel b). The star-formation episodes closely follow the cycles described in Section~\ref{sec:fbregulation} and Figure~\ref{fig:fbregulation}. As for the BH growth, we see no strong star formation inside the NSC during the starburst phase, consistent with the evolution of the nuclear gas supply in Figure~\ref{fig:fbregulation}. After the starburst phase, in-situ star formation grows the mass of NSC-1 (the cluster hosting BH-1) by $10^5 \, \rm M_\odot$ and that of NSC-2 (hosting BH-2) by $3 \times 10^4 \, \rm M_\odot$ before the NSC coalescence, and the mass of new stars reaches  ${\sim}2 \times 10^5 \, \rm M_\odot$ in the new star cluster formed by the NSC merger. Although there is a vast population of star clusters in the galaxy \citep{2020ApJ...891....2L}, we do not find evidence for the accretion of star clusters (aside from the merger of the two NSCs itself), and more than $95$ per cent of new stars that end up bound to the NSC at final time were formed bound to NSC-1, NSC-2, or the merged NSC.

Panel (c) shows that the half-mass radius grows from ${\sim} 3$ to ${\sim} 6 \, \rm pc$ during the NSC merger and continues to increase as stars are ejected by the BH binary. Finally, the BH recoil causes the NSC to puff up to $r_{1/2} \sim 10 \, \rm pc$. Similarly, the BH sphere of influence (radius enclosing $2 M_{\rm BH}$ in stars) expands as the central density decreases and the BH mass grows via gas accretion.

\subsubsection{Redistribution of NSC stars due to scattering and stripping}

Several processes redistribute mass from the nuclear star cluster to larger radii, inside and outside the galaxy: (gravitationally softened) star--star encounters, tidal stripping, and the unsoftened scattering by the BH binary.

To follow the stars that are ``lost'' from the NSC, i.e. stars that were bound to the NSC at some point but became unbound, we track the NSC membership of every star particle in the {\sc subfind} catalogues of the simulation snapshots. Before NSC coalescence, we combine the members of both clusters, so that stars exchanged between the two NSCs are not counted as lost. We count a particle as lost from the NSCs if it has been a member for five consecutive snapshots and is absent from the following five (${\sim} 18\,\rm Myr$ on either side). Because particles can leave and re-join the NSC several times, we assign each particle to the time of its last departure and record its position at $t = 860\,\rm Myr$ ($12\,\rm Myr$ after the BH merger). We consider a particle bound to the galaxy if its velocity does not exceed the escape velocity from the galaxy as computed from its spherically averaged potential.

Figure~\ref{fig:histogram} shows the result, split into particles that last left the NSC before the NSC coalescence ($0$--$500\,\rm Myr$, panel a), during the NSC coalescence ($500$--$508\,\rm Myr$, panel b), and during the BH binary scattering phase ($508$--$848\,\rm Myr$, panel c).

Before the NSC coalescence (panel a), the two NSCs together lose about $10^5\,\rm M_\odot$ over ${\sim} 500\,\rm Myr$, only ${\sim} 6$ per cent of their combined bound mass of $2 \times 8.2 \times 10^5\,\rm M_\odot$. Almost all of this mass stays within $1\,\rm kpc$ of the NSC and is still bound to the galaxy at the final time; about 30 per cent of it has even re-joined the NSC. During the coalescence of the NSCs (panel b), the NSC merger remnant loses $1.7\times10^5\,\rm M_\odot$, more than 10 per cent of the masses of the progenitor NSCs, within $8\,\rm Myr$. This mass is not ejected: it stays at a median distance of ${\sim} 80\,\rm pc$ from the NSC and remains bound to the galaxy. In the BH binary phase (panel c), the mass-loss rate is much lower than during the NSC coalescence, but the total lost mass is the largest of the three phases ($2.7\times10^5\,\rm M_\odot$), and much of it is scattered to large radii: about 40 per cent ends up beyond $1\,\rm kpc$, and about 13 per cent ($3.4\times10^4\,\rm M_\odot$) is unbound from the galaxy. These scattered stars are visible as outward-moving dots in Figure~\ref{fig:ejected_nsc_stars}. The total stellar mass lost during the BBH scattering phase is similar to the total mass of the coalescing BBH, which is consistent with the core-scouring picture that predicts the removal of stellar mass of the order of the BBH mass \citep{Merritt_2006}.

This re-distribution of mass across scales also has a strong effect on the density and surface brightness profiles of the NSC and the galaxy, which we discuss in Section~\ref{sec:discussion_observables}.

\subsection{Potentially observable signatures}
\label{sec:discussion_observables}

The simulations suggest several possible observational signatures of IMBH binary evolution in nucleated dwarf galaxies. We start by connecting the evolution of the NSC density profile to the observable surface brightness profiles, estimate the rate of TDEs from resolved encounters between stars and the BH, and make predictions for the gravitational-wave emission during the BBH merger.

\subsubsection{Mass-size relation and surface brightness profiles}
\label{sec:mock_photometry}

To compute surface-brightness profiles of the NSCs, we centre on the NSC
(Section~\ref{sec:merger_recoil}) and project along the $z$ axis. Stars
that formed during the simulation carry their individually sampled initial
mass, age and metallicity, and each of them is assigned the Hubble Space
Telescope (HST) ACS/WFC F814W luminosity of a single star of that initial
mass on the corresponding MIST v1.2 isochrone
\citep{Choi_2016,Dotter_2016}.
We assume that the pre-existing stars of the NSC and of the host galaxy
are old, with an age of $10\,\rm Gyr$, and assign their particles a
luminosity per unit mass from a simple stellar population (SSP) with
$[{\rm Fe}/{\rm H}]=-1$, giving $M/L_{\rm F814W}=1.6$ when stellar
remnants are counted in the mass. Light and stellar mass are summed in
logarithmic annuli, and we neglect dust, nebular emission and the point
spread function (PSF).

To estimate the size of the clusters, as in observations
\citep{Georgiev_2014,Neumayer_2020}, each profile is fitted between
$0.05$ and $2000\,\rm pc$ with the sum of an NSC and a host S\'ersic
component, with all six parameters free. We determine the effective 
radii of the NSC components of the light
and surface density fits, and $M_{\rm NSC}$ as the total mass of the
NSC component of the surface density fit. These masses agree with the bound
stellar mass found by {\sc subfind} to within $15\,$per cent.

In Figure~\ref{fig:nsc_obs}, we show the evolution of the cluster sizes,
inferred from the surface brightness profile (large circles) and the
surface density profiles (small circles) in comparison with the observed
NSCs of \citet[][NSCs only]{Norris_2014,Georgiev_2016}. At all times, the
properties of our simulated cluster are similar to the observed NSCs, even
though the stellar mass of the simulated galaxy lies at the low-mass end of
the host galaxies in the observational samples.

The surface-density profile (right) clearly shows evolution throughout the
stages of the merger. The NSC merger first raises the central density
($508\,\rm Myr$); scattering of stars by the BBH then reduces it again by
the time of the BH merger ($848\,\rm Myr$), and the ejection of the merger
remnant finally leaves a flat core of pc size \citep[see
also][]{Rawlings_2025}. Both effects combined lead to a reduction in central surface density of more than one order of magnitude. The same is visible in the surface brightness profile, although the reduction of the central density is less visible as a
function of time, because individual young, centrally concentrated stars
dominate the luminosity in the centre of the cluster. The change in the
profiles leads to an evolution of the mass and effective radius over time,
with $R_{\rm e} \sim 3 \, \rm pc$ inferred from the light profile
increasing to ${\sim} 6 \, \rm pc$ after the ejection of the BH.

\begin{figure}
 \includegraphics[width=\columnwidth]{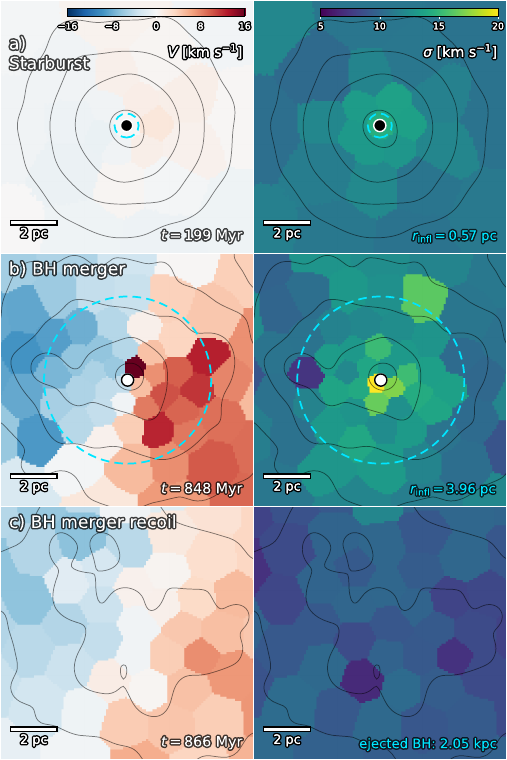}
\caption{Stellar kinematics of the NSC hosting BH-1 in
\simname{1e3-NSC}, projected along the $z$-axis. Rows: starburst phase at $199\,\rm Myr$ (a), BH merger (b) and
${\sim}18\,\rm Myr$ after the recoil kick (c). The left column shows the F814W light-weighted mean line-of-sight velocity $V$ and the right column the velocity dispersion $\sigma$, computed in Voronoi bins. Stars formed
in the simulation are weighted by the luminosity of their individually
sampled mass, age and metallicity (MIST isochrones), the pre-existing
population as a $10\,\rm Gyr$, $[\rm Fe/H]=-1$ stellar population. Thin
black contours are isophotes spaced by $1\,\rm mag$ (light smoothed over
$0.6\,\rm pc$). The dashed cyan circle marks the BH sphere of influence
$r_{\rm infl}$, $M_\star(<r_{\rm infl})=2M_{\rm BH}$, around the BH
(binary). Single bins with extreme $V$ or $\sigma$ in (b) and (c), and
the central $\sigma$ peak in (b), are dominated by the light of a few
luminous young stars. In (c) the merger remnant is $2.05\,\rm kpc$ from
the cluster.}
\label{fig:nsc_kin}
\end{figure}

The merger of the NSCs, the BBH, and its ejection also leave a kinematic
imprint on the cluster (Figure~\ref{fig:nsc_kin}, projected along $z$). At $199\,\rm Myr$,
during the starburst phase, the cluster hosting BH-1 shows no measurable
rotation, with a light-weighted spin parameter \citep{Emsellem_2007}
$\lambda_R(<4\,\rm pc)=0.06$. The NSC merger spins up the pre-existing
stars moderately \citep[$\lambda_R=0.24$, mass-weighted, at the time
of the BH merger; see also][]{Hartmann_2011,Tsatsi_2017}, whereas the stars
that form in situ around the binary inherit the angular momentum of the gas
and rotate faster ($\lambda_R\approx0.5$). Since they emit ${\sim}75$ per
cent of the F814W light within $4\,\rm pc$ while contributing only a
quarter of the mass, the light-weighted kinematics are those of a fast
rotator with $|V|\sim10-15\,\rm km\,s^{-1}$ and
$\lambda_R(<4\,\rm pc)=0.45$, comparable to the nuclei of late-type
galaxies, in which the rotation is likewise attributed to gas inflow
\citep{Pinna_2021}, and above those of early-type galaxy nuclei
\citep{Lyubenova_2019}. A few luminous young stars dominate individual
bins. Inside $r_{\rm infl}$, stellar scattering by the binary has removed
stars on radial orbits, which leaves a tangentially biased centre. The
intrinsic velocity anisotropy
$\beta = 1-(\sigma_\theta^2+\sigma_\phi^2)/(2\sigma_r^2)$, computed from
the mass-weighted velocity dispersions of all stars in spherical shells
with the mean rotation subtracted (not shown), is $\beta\sim-0.15$ at
$1$--$2\,\rm pc$. About $18\,\rm Myr$ after the recoil, the velocity
dispersion has dropped from ${\sim}13$ (outside the central spike
produced by a few young stars) to ${\sim}9\,\rm km\,s^{-1}$ and is flat
across the core, as expected from the virial response to the loss of
the BH and the expansion of the cluster. Rotation is weaker
($\lambda_R=0.30$) and the tangential bias has disappeared
($\beta=0\pm0.1$ at $1$--$3\,\rm pc$). The recoil-emptied NSC therefore
appears as a rotating cluster with a flat dispersion profile and no central
$\sigma$ rise. Velocity dispersions of ${\sim} 10\,\rm km\,s^{-1}$ within a
parsec-sized core are at the limit of what current instruments can measure,
but instruments such as HARMONI on the Extremely Large Telescope (ELT)
\citep{Thatte_2021} could measure the rotation and the flat dispersion
profile of such a core, from which dynamical modelling could constrain its
velocity anisotropy and the absence of a central BH.

\subsubsection{Tidal disruption events}

Tidal disruption events of stars (TDE) are one of the most promising tools for detecting non-accreting IMBHs \citep[][see \citealt{Gezari_2021} for a review]{Lin_2018, Greene_2020}. While our simulations do not follow the full collisional $N$-body dynamics, they are in a unique position to trace the collisional interaction between individually sampled new stars and IMBHs on the fly \citep[see also][]{2025MNRAS.537..956P,Lahen_2025b}. In galaxy-scale simulations that do not resolve the stellar population around the BH, TDE rates have instead to be modelled with a subgrid prescription for the loss-cone dynamics, as done by \citet{Pfister_2021} in a cosmological simulation. We do not allow TDEs as a growth channel in the simulation, but track each encounter that is below the tidal disruption radius \citep[as implemented in][]{2025MNRAS.537..956P}

\begin{equation}
 r_{\rm t}
    = 1.3\,R_\star
 \left(
 \frac{M_{\rm BH}+m_\star}{2m_\star}
 \right)^{1/3},
\end{equation}
assuming a main sequence stellar radius of
\begin{equation}
 R_\star
    = R_{\odot}
 \left(
 \frac{m_\star}{M_\odot}
 \right)^{0.8}.
\end{equation}

\begin{figure}
	\includegraphics[width=1\columnwidth]{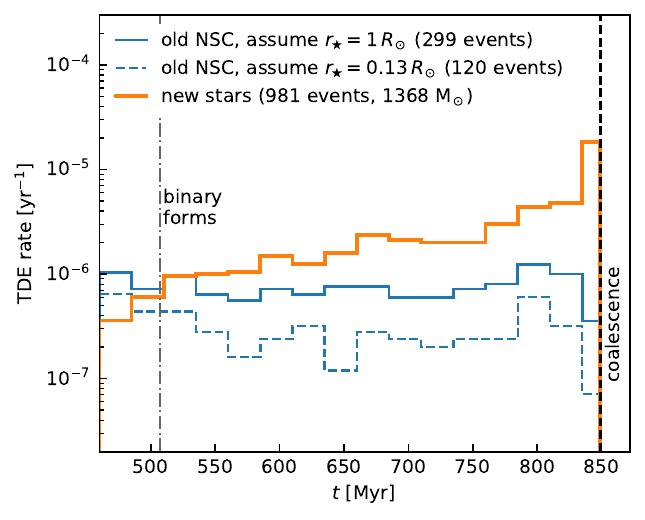}
 \caption{Rate of tidal disruptions by the two nuclear BHs in the \simname{1e3-NSC}
 run, in $25\,\rm Myr$ bins, from $t\simeq460\,\rm Myr$
 to the coalescence of the BH binary at $t=848\,\rm Myr$ (black dashed line). The
 dot-dashed line marks the formation of the bound binary at $t\sim508\,\rm Myr$. A
 particle is counted as a TDE when it crosses the tidal radius $r_{\rm p}<r_{\rm t}$ for the first time.
 Orange: all individually sampled new star particles are evaluated at each particle's own mass. Blue: the $20\,\rm M_\odot$ pre-existing NSC particles, which do
 not represent single stars. The solid and dashed histograms show the same particles
 under an assumed $R_\star=1\,\rm R_\odot$ and $0.13\,\rm R_\odot$ (i.e.\ $1$ and
 $0.08\,\rm M_\odot$ on the same mass--radius relation).}
 \label{fig:tde_rate}
\end{figure}

This allows us to reconstruct the number of encounters with pericentre passages below the disruption radius in postprocessing. Taking into account that stars can only be disrupted once, we find that 981 new stars crossed the tidal radius between $460$--$848\,\rm Myr$, representing ${\sim} 1368 \, \rm M_\odot$ disrupted stellar mass. In comparison to the accreted gas mass of $\sim 8 \times 10^4 \, \rm M_\odot$ during the same period of time, TDEs of new stars would hence contribute $\sim 1$ per cent to the growth of the BHs, at the lower end of the 1--10 per cent predicted for IMBHs by \citet{Polkas2024}. As the TDE rate as a function of time in Figure~\ref{fig:tde_rate} shows, the rate increases steeply towards the IMBH coalescence. This differs from models of a binary in a fixed stellar cusp, in which the TDE rate is enhanced shortly after binary formation and then declines as the loss cone is depleted \citep{Chen_2009,Chen_2011}, possibly due to a combination of the effects of ongoing in-situ star formation inside the NSC and numerical relaxation effects, that might artificially replenish the stars available for disruption (see Section \ref{sec:discussion_caveats_dynamics}). 

As old NSC stars are not sampled as individual stars but have a uniform mass of $20 \, \rm M_\odot$, we can only make an order-of-magnitude estimate for their TDE contribution. Assuming the tidal disruption criterion of a $0.08\,\rm M_\odot$ star ($R_\star = 0.13\,\rm R_\odot$), we find 120 events. If we assume a larger stellar mass ($1\,\rm M_\odot$, $R_\star = 1\,\rm R_\odot$), the  number of disruption events rises to ${\sim}$300. Because in this approximation each star particle that fulfils the disruption criterion now represents multiple individual stars, their TDE rate will likely be higher, but we do not attempt to model this here.

\subsubsection{Runaway and hyper-velocity stars}

Our simulations produce a population of runaway and hyper-velocity stars through interactions with the BHs and the central binary. Such stars may provide an indirect signature of a massive BH binary \citep[e.g.][]{Sesana_2006,Cao2025}. In addition, runaway stars that transport SN explosions away from their birth sites can alter how feedback couples to the surrounding ISM \citep{Steinwandel2023}. We find stars with a total mass of ${\sim} 3.4 \times 10^4 \, \rm M_\odot$ with velocities between $300$ and $2000 \, \rm km\,s^{-1}$ that become unbound from the galaxy. However, the delay between nuclear star formation and the scattering of these stars typically exceeds the stellar evolution timescale of massive stars in our simulations, and only eight ejected stars more massive than $8\,{\rm M_\odot}$ are still alive at the time of ejection, of which three subsequently explode more than ${\sim}1\,{\rm kpc}$ away from the NSC.

\subsubsection{Gravitational-wave signatures}
\label{sec:gw_signatures}

The IMBH coalescences in the simulations \simname{1e3-NSC} and \simname{1e3-NSC-reset} (Section~\ref{sec:Coalesence}) fall into the mass range that is expected to be detectable by the future gravitational-wave observatory Laser Interferometer Space Antenna (LISA) \citep{AmaroSeoane_2017,Colpi_2024}. To estimate the observational signature explicitly, we compute the characteristic strain of the coalescing binaries following the prescription of \citet{Keitaanranta2026}, who also use {\sc Ketju}, albeit in a different BH mass range.

The simulation provides the semimajor axis and eccentricity of the binary down to its last output, at $a \sim 100\,R_\mathrm{S}$ for the fiducial binary, and we continue the orbit to coalescence with the orbit-averaged equations of \citet{Peters_1964}. We evaluate the characteristic strain of the first 300 harmonics of the orbital frequency \citep{PetersMathews1963} along the orbit, limit the number of wave cycles to those accumulated during a mission of 4~yr (in the observer frame), and sum the harmonics in quadrature. Above twice the frequency at which this inspiral spectrum peaks, where the orbit has circularised to $e < 10^{-3}$, we use instead the non-spinning IMRPhenomD waveform \citep{Husa2016,Khan2016} for the late inspiral, merger and ringdown (the two descriptions agree to about 2 per cent at the transition). All strains are averaged over sky position, inclination and polarisation \citep{FinnThorne2000}. The signal-to-noise ratio (SNR) follows from integrating the squared ratio of the characteristic strain to the LISA noise amplitude of \citet{Robson2019}, including the Galactic foreground for a 4-yr mission, over logarithmic frequency.

\begin{figure}
 \centering
 \includegraphics[width=\columnwidth]{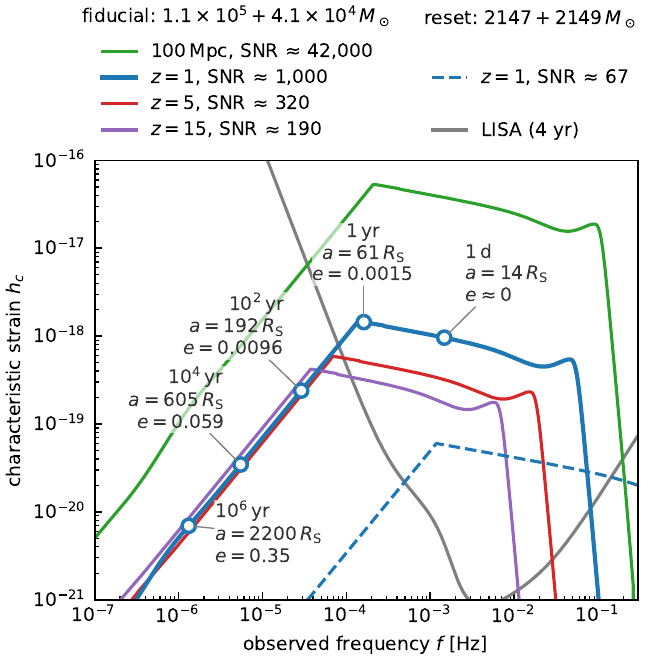}
 \caption{Characteristic strain of the fiducial IMBH binary
 (\simname{1e3-NSC}, $1.1\times10^5 + 4.1\times10^4 \, \rm M_\odot$, solid) if the simulation is placed at $z\sim0$ (100~Mpc),
    1, 5, and 15, and of the $2.1\times10^3 + 2.1\times10^3 \,\rm M_\odot$ binary
 of the reset run (\simname{1e3-NSC-reset}, dashed, $z=1$), compared with the LISA sensitivity
 curve of \citet{Robson2019} for a four-year mission. Curves combine the
 simulated orbit, its continuation using the Peters formalism,
 and the IMRPhenomD merger--ringdown. The legend gives the sky-averaged
 SNR. Markers on the fiducial $z=1$ curve indicate the observed time before merger,
 the binary semimajor axis in $R_{\rm S}=2GM_{\rm tot}/c^2$, and the
 eccentricity.}
 \label{fig:LISA}
\end{figure}

Figure~\ref{fig:LISA} shows the resulting characteristic strain at several redshifts \citep[assuming cosmological parameters from][]{Planck2016} and at 100~Mpc. For the fiducial binary the sky-averaged SNR reaches approximately $1000$ at $z=1$, $320$ at $z=5$, $190$ at $z=15$, and $4\times10^4$ at 100~Mpc, showing that the merger in the fiducial simulation would be detectable regardless of the redshift at which the (redshift-agnostic) simulation is placed. On the other hand, the lighter binary in the reset run has a weaker signature, but is still detected with an SNR of $\simeq67$ at $z=1$. For these sources, LISA might even be able to detect a residual eccentricity \citep{Garg2024}, but we do not attempt to model this here.

\subsubsection{AGN luminosity}\label{sec:agn_luminosity}
Because we deliberately do not model unresolved accretion-disc processes or AGN feedback, both of which can be important for regulating BH accretion in dwarf galaxies (Petersson et al., in preparation), we cannot give detailed predictions for the X-ray luminosities of the coalescing BHs. However, the accretion history in Figure~\ref{fig:fbregulation} provides an upper limit on BH growth and shows that the BHs are inactive most of the time, and hence invisible to techniques that rely on active accretion, such as X-ray surveys. In particular, our simulations predict no AGN activity during the starburst, when the galaxy itself would be brightest. Before and after the starburst, accretion is enhanced compared to an isolated galaxy, qualitatively consistent with observations of enhanced AGN fractions in dwarf galaxy pairs \citep{Micic_2024,Erostegui_2025}. In the post-starburst phase, BH activity occurs in brief episodes that reach or exceed the Eddington rate for a few Myr (Figure~\ref{fig:fbregulation}, where the accretion history is smoothed over $2\,\rm Myr$). These episodes correspond to bolometric luminosities of order $10^{43}\,\rm erg\,s^{-1}$, or X-ray luminosities of up to ${\sim} 10^{42}\,\rm erg\,s^{-1}$ for typical bolometric corrections; in the absence of AGN feedback, these values are upper limits. During such episodes, a BH with the accretion history found here would be comparable to the most luminous X-ray-selected AGN in dwarf galaxies \citep{Bykov_2024,Sacchi_2024}, but with a duty cycle of only ${\sim} 10$ per cent, so that even in the post-starburst phase the majority of such IMBHs would likely be missed by observational searches.

\section{Discussion}
\label{sec:Discussion}

 In the following, we interpret the main physical mechanisms controlling BH coalescence and growth, discuss the principal limitations of our modelling, and consider the implications for hierarchical IMBH growth.

\subsection{Nuclear star clusters as catalysts of IMBH coalescence}
\label{sec:discussion_nsc_dynamics}

The inefficient sinking of low-mass IMBHs without an accompanying NSC is consistent with previous work showing that light BHs can remain on long-lived off-centre orbits \citep{Pfister_2019,Bellovary_2021,Ma_2021,2024MNRAS.532.4681P,Zhou_2025}. Embedding the BH in an NSC increases the effective dynamical mass of the sinking object and therefore accelerates dynamical-friction-driven orbital decay \citep[e.g.][]{Mukherjee_2025}. At smaller separations, however, the extended nature of the stellar cluster also becomes important. \citet{Ogiya_2020} showed that tidal stripping during the interaction of two NSCs can provide an additional channel for orbital-energy loss, accelerating their coalescence. We find a similar behaviour in the simulations presented here. In particular, replacing the NSC by a BH of equivalent mass does not reproduce the same rapid evolution towards a bound BH binary, consistent with the related experiment of \citet{Mukherjee_2025}. These results suggest that the NSC affects BH pairing not only through its additional mass, but also through the tidal evolution of the extended stellar nuclei.

Once a hard binary has formed, the limiting process changes from transporting the BHs to maintaining a sufficiently dense stellar reservoir for continued three-body scattering \citep{Quinlan_1996,Sesana_2007,Vasiliev_2015,Khan_2024}. In particular, \citet{Khan_2021} find that the hardening efficiency depends strongly on the stellar density around the binary, which is expected to decrease as the binary becomes increasingly massive relative to its environment. Their models further show that the NSC dominates the supply of scatterers when $M_{\rm NSC}/M_{\rm BBH}\gtrsim10$, whereas stars from the surrounding bulge contribute increasingly at lower NSC-to-binary mass ratios. This provides a useful framework for interpreting our fiducial and mass-reset calculations: a binary that becomes dynamically important relative to its host cluster more strongly perturbs and depletes the central stellar distribution, whereas a lighter binary can continue to harden without substantially modifying the surrounding NSC. This is also consistent with the classical core-scouring picture, in which a hard BH binary ejects a stellar mass of order its own mass while shrinking \citep{Milosavljevic_2001,Milosavljevic_2003,Merritt_2006}.

In our simulations, the pre-existing NSC remains the dominant stellar reservoir during the hardening phase (Section~\ref{sec:Coalesence}). Taken together, these results suggest that the relevant quantity for IMBH coalescence is not the BH mass alone, but the evolving relation between the binary mass and the density and structure of the stellar reservoir available for hardening. Increasing the BH mass can therefore accelerate the initial sinking while making the later stellar-hardening phase less efficient if the binary becomes too massive relative to its host cluster. The simulations therefore suggest that there might be an optimal mass ratio that leads to the fastest IMBH coalescence, but a systematic parameter study would be required to study this effect.

\subsection{Stellar feedback-regulated BH growth during the galaxy merger}
\label{sec:discussion_bh_growth}

Even in the gas-rich environment of a galaxy merger, IMBHs without an accompanying NSC essentially fail to grow through gas accretion, consistent with idealised simulations of IMBHs in isolated dwarf galaxies \citep{2025MNRAS.537..956P,2025A&A...704A.177P,2025MNRAS.544.2467S}. The presence of an NSC alone, however, is not sufficient to guarantee efficient growth. During the merger-induced starburst at ${\sim}150$--$300\,{\rm Myr}$, we find little BH growth even in the fiducial NSC model despite the high global gas densities, because the high SN rate and the streaming velocity of the NSCs relative to the gas prevent a persistent nuclear gas reservoir (Section~\ref{sec:fbregulation}).

Once the NSCs settle and the starburst subsides, the system transitions to a locally regulated regime (Section~\ref{sec:fbregulation}). Similar feedback-regulated accretion cycles have been found in isolated dwarf-galaxy simulations \citep{2025MNRAS.537..956P,2025MNRAS.544.2467S,Shin2026}. The orbital evolution contributes as well: as the BH--NSC systems sink, they spend progressively more time at lower relative velocity with respect to the dense gas in the central regions of the merger remnant, where repeated gas capture becomes more favourable. Prior to the starburst, growth is also enhanced compared to the isolated galaxy (Section~\ref{subsec:growth}).

The suppression of accretion during the starburst is broadly consistent with the supernova-regulated BH growth found in lower-resolution cosmological simulations of low-mass galaxies \citep{Dubois_2015,AnglesAlcazar_2017,Habouzit_2017,Trebitsch_2018,Koudmani_2022}. Our resolved ISM allows us to distinguish two regimes more directly: an early starburst phase in which gas capture is globally suppressed, followed by a locally regulated phase in which orbital evolution and stellar feedback together control the nuclear gas supply. This distinction also highlights the importance of resolving the orbital evolution of the BHs. Artificial repositioning, or merging the BHs before their orbital decay is resolved, could place them prematurely in conditions favourable for gas capture and therefore overestimate their growth.

\subsection{Numerical limitations and unresolved physics}
\label{sec:discussion_caveats}

Our quantitative merger times and BH growth histories should be interpreted in light of several numerical and physical limitations. These primarily affect the detailed hardening rates, exact merger times, and BH masses. At the same time, our conclusions that gas capture is suppressed during the starburst but enhanced relative to an isolated galaxy in the post-starburst phase, and that an NSC can enable IMBH pairing, ensure continuous stellar hardening, and lead to a BH merger are less sensitive to them. 

\subsubsection{Stellar dynamics and numerical resolution}
\label{sec:discussion_caveats_dynamics}

With our choice to follow the unsoftened BH--star encounters but to use softened star--star dynamics, we are able to simulate the evolution of a massive NSC in its full galactic environment, but miss several processes that can only be fully resolved by future studies employing direct $N$-body codes with individual stars.

For example, our modelling choice affects the two-body relaxation time \citep{Spitzer_1987}, because the star--star softening removes encounters with impact parameters below $\epsilon_{\rm soft} = 0.05\,\rm pc$, which suppresses relaxation. At the same time, the uniform mass resolution of $20 \, \rm M_\odot$ of the old NSC population shortens the relaxation time. This possibly affects how the inner density profile of the cluster evolves and how the loss cone of the binary is refilled, such that the exact hardening rate should likely be regarded as an upper limit. We note that this limitation is common to simulations of IMBH binaries in dense stellar systems and affects most of them to varying degrees. For example, the $N$-body studies that report efficient hardening in NSCs used equal-mass stellar particles of ${\sim} 250$--$3000\,\rm M_\odot$ \citep{Khan_2021} and ${\sim} 60$--$500\,\rm M_\odot$ \citep{Mukherjee_2025}, i.e. they are more collisional than our NSC model, whereas \citet{Khan_2026} have recently reached a resolution of $1 \, \rm M_\odot$, at which the relaxation time is close to its physical value. The same authors find that a mass-segregated population of stellar-mass BHs can nearly double the sinking rate and reduce the merger time by about a factor of two. Such a population is not represented self-consistently in our pre-existing NSCs, which consist of stellar particles of uniform mass.

For the fiducial model, the BH-to-NSC-particle mass ratio during binary hardening satisfies $M_{\rm BH}/m_{\rm NSC}\gtrsim10^3$, within the regime in which \citet{Mannerkoski_2023} find converged hardening rates in {\sc Ketju} tests. The mass-reset experiment is less well resolved in this respect because the masses of individual massive stars and of NSC particles become non-negligible relative to the BH masses ($M_{\rm BH}/m_{\rm NSC}\gtrsim 10^2$), and its exact hardening timescale should therefore be interpreted more cautiously. More generally, the binary eccentricity and hence the precise coalescence time can be sensitive to stochastic stellar encounters and to the orbital evolution preceding binary formation \citep{Nasim_2020,Rawlings_2023}. However, this is not purely a numerical effect, as the mass of BHs at the lower end of the IMBH mass range becomes comparable to the physical mass of individual stars, such that their coalescence will depend more directly on the properties of the stellar mass function and physically resolved strong encounters that abruptly change the orbital evolution. We also note that the BH merger in the fiducial simulation does not rely on an initially extreme eccentricity, as the binary forms with $e = 0.7$. Only after the formation of a hard binary does stellar scattering slowly drive up the eccentricity to $e = 0.96$.

Previous {\sc Ketju} tests and our earlier resolution study indicate that dynamical-friction-driven sinking is substantially less sensitive to particle mass resolution than binary hardening \citep{Mannerkoski_2023,2024MNRAS.532.4681P,Keitaanranta2026}. Nevertheless, the slowly sinking low-mass no-NSC systems should be interpreted with the finite resolution of the dark-matter background in mind \citep{Genina2024} ($M_{\rm BH}/m_{\rm DM} = 2.5$ and $25$ for the $10^3$ and $10^4\,\rm M_\odot$ BHs, respectively).

\subsubsection{Unresolved gas physics and BH feedback}

Because of our finite spatial resolution and adopted accretion radius, we do not resolve the interaction of the BH binary with gas below ${\sim}0.5\,{\rm pc}$. Hydrodynamical simulations of more massive BH binaries show that gas can continue to contribute to orbital evolution on sub-parsec scales, although both the efficiency and the direction of the resulting torques depend sensitively on the local gas density, thermodynamics, and disc structure \citep{Escala_2005,Dotti_2007,Mayer_2007,SouzaLima_2020}. It is therefore unclear whether unresolved gas torques would significantly alter the merger times in the much lower-mass IMBH systems considered here. 

On smaller scales, circumbinary accretion can modify both the binary mass ratio and spin configuration. Simulations generally find preferential accretion onto the lower-mass BH, tending to drive the mass ratio towards unity \citep{Duffell_2020,Munoz_2020,Siwek_2023b}. In this study, we do not model this differential accretion or spin evolution, and the gravitational-wave recoil velocities inferred here should therefore be regarded as uncertain, since they depend sensitively on both quantities \citep{2015PhRvD..92b4022Z}. We also note that it is unclear how dynamically important gas effects are in this set-up, as the gas mass is generally subdominant compared to the stellar mass in the vicinity of the BH, and never exceeds ${\sim}300 \, \rm M_\odot$.  

Finally, we do not include BH feedback. Previous simulations show that feedback can suppress nuclear gas inflows and star formation \citep{Liao_2024a,Liao_2024b} and evacuate the gas around sinking BHs, weakening dynamical friction \citep{SouzaLima_2017}, and thereby alter BH pairing and stellar hardening. In our simulations, the pre-existing NSC dominates the stellar reservoir, but feedback could still affect the coalescence indirectly by regulating BH growth and hence the evolving BBH-to-NSC mass ratio. Preliminary simulations including BH wind feedback indicate substantially reduced BH growth, suggesting that the growth of the IMBHs in the fiducial \simname{1e3-NSC} simulation should be regarded as an approximate upper limit.

\subsubsection{Generality of the merger timescale}

Our merger times should not be interpreted as universal for nucleated dwarf galaxies. We follow a single, computationally expensive, high-resolution merger configuration with relatively massive and compact NSCs. Both NSC-assisted sinking and subsequent stellar hardening are expected to depend on the NSC mass, density and orbital structure, as well as on the galaxy merger orbit \citep{Khan_2021,MastrobuonoBattisti_2025,Mukherjee_2025,Wright_2026}.

\subsection{Implications for hierarchical IMBH growth}
\label{sec:discussion_hierarchical_growth}

Although our initial conditions are idealised, the simulations demonstrate that IMBH mergers in low-mass galaxies can occur on sub-Gyr timescales when the BHs are embedded in sufficiently dense NSCs. If light BH seeds \citep{Volonteri_2021} with masses of $\lesssim 1000 \, \rm M_\odot$ commonly reside in NSCs, the inefficient large-scale sinking that is expected for these light BHs in isolation does not by itself prevent hierarchical BH growth in clumpy galaxies at high redshift. However, the ${\sim}340\,{\rm Myr}$ delay between NSC coalescence and BH merger further shows that unresolved BH dynamics can introduce substantial delays between galaxy and BH mergers, even if embedded in an NSC, which should be accounted for in cosmological models of BH growth and in predictions of BH merger rates.

Efficient coalescence does not, however, imply efficient BH growth or retention of the BH inside the host cluster. The BH growth history depends on the interplay between orbital evolution, stellar feedback, and nuclear gas capture, while gravitational-wave recoil can remove the merger remnant altogether. In our fiducial model, the recoil ejects the remnant from the NSC, leaving behind a stellar nucleus without a central BH and a wandering black hole in the galactic outskirts. Hence, merger-and-ejection events of this kind could influence the BH occupation fraction \citep{Greene_2020} and the fraction of off-centre BHs in present-day dwarf galaxies \citep{Reines_2020}, and produce wandering massive BHs like the off-nuclear candidate recently reported by \citet{Patra_2025} and \citet{Yao_2025}.

Taken together, these results suggest that hierarchical IMBH growth depends on several coupled processes: delivering the BHs to the nucleus, maintaining a stellar reservoir for binary hardening, growing the BHs before coalescence (but not too much to deplete the NSC), and retaining the merger remnant. NSCs can substantially alleviate the first two bottlenecks, but the last two remain sensitive to feedback and gravitational-wave recoil.

A comprehensive assessment of this pathway will require a broader range of galaxy structures, NSC properties, and merger orbits, ideally drawn from a cosmological environment. Such calculations are currently difficult at the sub-parsec resolution required to resolve both BH dynamics and gas capture by the NSCs. The simulations presented here therefore provide a resolved case study of the physical processes that need to be captured in models of hierarchical IMBH growth.

\section{Summary and Conclusions}
\label{Sec:summary}

We have presented a simulation of a gas-rich merger of two equal-mass dwarf galaxies, each hosting a $10^3\,\rm M_\odot$ IMBH embedded in a $10^6\,\rm M_\odot$ NSC, together with comparison runs without NSCs. The simulations combine the resolved multiphase ISM, individually sampled stars and sink-particle accretion of {\sc griffin} with the regularised post-Newtonian dynamics of {\sc Ketju}, and follow the clusters and black holes from the first pericentre passage of the galaxies to the gravitational-wave-driven merger of the IMBHs. The central result is that NSCs and IMBHs co-evolve: the clusters enable the pairing, growth and merger of the black holes, and the binary and its merger in turn reshape the cluster. We summarise the main findings along the four stages of this pathway and its observable signatures.

\begin{itemize}

\item \textit{Delivery and pairing.} Without NSCs, the $10^3$ and $10^4\,\rm M_\odot$ IMBHs sink only to separations of $50$--$200\,\rm pc$ within $900\,\rm Myr$ and do not form a bound pair, which a rough Chandrasekhar estimate predicts would not happen before ${\sim} 2\,\rm Gyr$ after the end of the simulation for the $10^4\,\rm M_\odot$ BHs, and not within a Hubble time for the $10^3\,\rm M_\odot$ BHs. With NSCs, the clusters initially sink as $10^6\,\rm M_\odot$ bodies, overlap at pericentre separations of $\lesssim 10\,\rm pc$ and coalesce at ${\sim} 508\,\rm Myr$, losing ${\sim} 10$ per cent of their mass to tidal stripping. Inside the merged cluster, the IMBHs sink to ${\sim} 0.1\,\rm pc$ and form a binary within a few Myr.

\item \textit{Growth before coalescence.} The NSCs promote nuclear gas capture and grow the sink particles to $1.1\times10^5$ and $4.1\times10^4\,\rm M_\odot$, whereas IMBHs without NSCs capture negligible gas. Compared with the same galaxy evolved in isolation, where the IMBH grows to only ${\sim} 1.3\times10^4\,\rm M_\odot$ over the same period, the merger enhances BH growth by about an order of magnitude. Capture is inefficient during the merger-induced starburst at ${\sim} 150$--$300\,\rm Myr$, despite the high gas density, and begins only after the global SN rate has declined and the NSCs have settled in the new galactic centre, in brief episodes regulated by nuclear star formation and local SNe. Without accretion-disc physics and AGN feedback, these masses are upper limits.

\item \textit{Hardening and coalescence.} Scattering of cluster stars drives the binary into the gravitational-wave regime, and the IMBHs merge at $t_{\rm merge}\sim 848\,\rm Myr$, about $340\,\rm Myr$ after the clusters coalesce and less than $1\,\rm Gyr$ after the first pericentre passage of the galaxies. The merger does not rely on an initially extreme eccentricity: the binary forms with $e=0.7$ and is driven to $e=0.96$ by stellar scattering. Pre-existing NSC stars dominate the scattering, new stars contribute ${\sim} 20$ per cent of the ejected mass, and ongoing nuclear star formation causes no identifiable short-term increase of the hardening rate. The binary-to-NSC mass ratio controls both the hardening rate and the response of the cluster: resetting the IMBH masses to $10^3\,\rm M_\odot$ before NSC coalescence yields a lighter binary that perturbs the cluster less and merges earlier.

\item \textit{Retention and imprint on the NSC.} For the non-spinning BHs assumed here, the fiducial merger with $q=0.38$ produces a recoil of ${\sim} 175\,\rm km\,s^{-1}$ that ejects the remnant from the NSC but leaves it bound to the galaxy, whereas the near-equal-mass remnant of the reset run stays in the cluster. Scouring by the binary and the removal of the remnant flatten the central density profile and expand the cluster. In mock HST photometry, the effective radius grows from ${\sim} 3$ to ${\sim} 6\,\rm pc$ and the central surface density drops by more than an order of magnitude, while the cluster stays on the observed NSC mass--size relation. The merged NSC is a fast rotator ($\lambda_R\sim 0.45$ within $4\,\rm pc$) and, after the recoil, has a flat velocity-dispersion profile of ${\sim} 9\,\rm km\,s^{-1}$ without a central rise.

\item \textit{Observable signatures.} The hardening binary produces a TDE rate that rises steeply towards coalescence and ejects ${\sim} 3.4\times10^4\,\rm M_\odot$ of stars at $300$--$2000\,\rm km\,s^{-1}$ from the galaxy. The IMBHs are inactive most of the time. No AGN activity is predicted during the starburst, and the brief episodes of near- or super-Eddington accretion outside it have a duty cycle of only ${\sim} 10$ per cent, so that most such IMBHs would be missed by X-ray searches. The merger itself is a LISA source with a sky-averaged signal-to-noise ratio of ${\sim} 10^3$ at $z=1$ and ${\sim} 190$ at $z=15$.

\end{itemize}

We conclude that NSCs remove the two bottlenecks that most often stall IMBH mergers in dwarf galaxies, the delivery of the black holes to the nucleus and the supply of stars for binary hardening, and thereby allow a black hole merger within $1\,\rm Gyr$ of the galaxy merger. The NSC does not, however, guarantee efficient black hole growth or the retention of the merger remnant. Growth is confined to brief, stellar-feedback-regulated episodes after the starburst, and the gravitational-wave recoil can remove the merger remnant from the cluster. In our simulation, this mechanism leaves behind a nuclear star cluster without a BH, but with an IMBH wandering in the galaxy outskirts.

Several of the signatures found here will become testable in the coming decade. Wide-field time-domain surveys such as the Legacy Survey of Space and Time of the Vera C. Rubin Observatory \citep{Ivezic_2019} will detect large samples of TDEs, including off-nuclear events from recoiling or still-sinking black holes in dwarf galaxies, and an enhanced TDE rate would be a signature of a BH binary. Integral-field spectrographs on 30-m class telescopes, such as HARMONI on the ELT \citep{Thatte_2021}, will resolve the kinematics of nearby nucleated dwarfs on the parsec scales required to detect the rotation, flat dispersion profiles and cores left behind by a binary and its recoil. LISA \citep{Colpi_2024} will detect mergers of $10^4$--$10^5\,\rm M_\odot$ black holes like the one simulated here across most of cosmic history, so that the merger rate of IMBHs in dwarf galaxies, and with it the efficiency of the NSC-assisted pathway, will become directly measurable. To make detailed predictions for these observations, future simulations should take into account the full complexity of the coupled BH--NSC evolution, including the crucial effects of the resolved BH pairing and hardening, as well as the BH interaction with nuclear star formation facilitated by the highly structured ISM.

\section*{Acknowledgements}
We thank Rory Smith, Sara Ellison, Mudit Garg, Thorsten Naab, Jonathan Petersson, Peter Johansson, Jonathan J\"ager, Alexander Rawlings, Tiziana Di Matteo, and Debora Sijacki for valuable discussions. Computations were performed on the HPC system Rusty of the Simons Foundation/Flatiron Institute. The Flatiron Institute is supported by the Simons Foundation. CP and MH acknowledge funding from the Swiss National Science Foundation (SNSF) via a PRIMA grant PR00P2\_193577. Generative AI (OpenAI ChatGPT and Anthropic Claude) was used to assist with literature research, to improve the language and readability of this manuscript, and to assist in the modelling of observables in Sections~\ref{sec:mock_photometry} and \ref{sec:gw_signatures}.

\section*{Data Availability}

The data underlying this article will be shared on reasonable request to the corresponding author.



\bibliographystyle{mnras}
\bibliography{literature} 



\bsp	
\label{lastpage}
\end{document}